\documentclass[11pt]{article} % pas draft
\usepackage{amsmath}
\usepackage{amssymb,amsfonts,euscript}
\usepackage{hyperref}
\renewcommand{\baselinestretch}{1.2}
\newcommand\nn{\nonumber}

\newcommand{\A}{{\cal A}}
\newcommand{\Ord}{{\cal{O}}}
\newcommand{\Ozw}{\Ord (z-w)^0}
\newcommand{\C}{{\cal C}}

\newcommand{\G}{{\cal G}}

\newcommand{\T}{{\cal T}}

\newcommand{\F}{{\cal F}}

\newcommand{\lr}{{\cal{L}}_{\sgb(\s)} }

\newcommand{\Zm}{\mathbb{Z}}
\newcommand{\Cm}{\mathbb{C}}
\newcommand{\Cmg}{\{\mathbb{A}(2n;r)\}}
\newcommand{\Cmgi}{\mathbb{A}(2n;r)}
\newcommand{\Pm}{\mathbb{P}}
\newcommand{\Am}{\mathbb{A}}
\newcommand{\Tmm}{\mathbb{T}}
\newcommand{\Tm}{\mathbb{T}_1}
\newcommand{\Tmt}{\mathbb{T}_2}
\newcommand{\Cmo}{\frac{\so(2n)}{\mathbb{C}}}
\newcommand{\Cmgo}{\left\{\frac{\so(2n)}{\Cmgi}\right\}}

\newcommand{\Pmo}{\frac{\so(2n)}{\mathbb{P}}}
\newcommand{\Tmo}{\frac{\so(8)}{\mathbb{T}_1}}
\newcommand{\Tmto}{\frac{\so(8)}{\mathbb{T}_2}}
\newcommand{\Tmos}{\frac{\so(8)}{\mathbb{T}_1^2}}
\newcommand{\Tmtos}{\frac{\so(8)}{\mathbb{T}_2^2}}
\newcommand{\Tmmo}{\frac{\so(8)}{\mathbb{T}}}
\newcommand{\Tmmos}{\frac{\so(8)}{\mathbb{T}^2}}

\newcommand{\bz}{\bar{z}}

\def\a{\alpha}
\def\be{\beta}
\def\hb{\hat{\be}}

\def\m{\mu}
\def\n{\nu}
\def\r{\rho}
\def\de{\delta}
\def\k{\kappa}

\def\nl{\newline}

\def\ni{\noindent}

\def\hg{\hat{g}}
\def\hgp{\hat{g}_+}
\def\hgm{\hat{g}_-}
\def\ga{\gamma}
\def\part{\partial}
\renewcommand{\sp}{,\hspace{.3in}}
\def\o{\omega}
\def\gm{\mathfrak{g}}
\def\su{\mathfrak{su}}
\def\so{\mathfrak{so}}
\def\spi{\mathfrak{spin}}
\def\gt{\mathfrak{g}_2}

\newcommand{\sa}{\mathop{\vtop{\ialign{##\crcr
  $\hfil\displaystyle{\longrightarrow}\hfil$\crcr\noalign{\kern-1pt\nointerlineskip}
  \hspace{.12in}$^\sigma$\hskip6pt\crcr\noalign{\kern3pt}}}}}
\newcommand{\slra}{\mathop{\vtop{\ialign{##\crcr
  $\hfil\displaystyle{\longleftrightarrow}\hfil$\crcr\noalign{\kern-1pt\nointerlineskip}
  \hspace{.12in}$^\sigma$\hskip6pt\crcr\noalign{\kern3pt}}}}}
\newcommand{\sat}{\mathop{\vtop{\ialign{##\crcr
  $\hfil\displaystyle{\longrightarrow}\hfil$\crcr\noalign{\kern-1pt\nointerlineskip}
  \hspace{.12in}$^\sigma$\hskip6pt\crcr\noalign{\kern3pt}}}}}
\newcommand{\pa}{\mathop{\vtop{\ialign{##\crcr
  $\hfil\displaystyle{\oplus}\hfil$\crcr\noalign{\kern+1pt\nointerlineskip}
  \hspace{.08in}$^{\alpha=0}$\hskip6pt\crcr\noalign{\kern3pt}}}}}
\newcommand{\ka}{\mathop{\vtop{\ialign{##\crcr
  $\hfil\displaystyle{\longleftrightarrow}\hfil$\crcr\noalign{\kern-1pt\nointerlineskip}
  \hspace{.12in}$^K$\hskip6pt\crcr\noalign{\kern3pt}}}}}
\newcommand{\bp}{\mathop{\vtop{ialign{##\crcr
  $\hfil\displaystyle{}\hfil$\crcr\noalign{\kern-13pt\nointerlineskip}
  \big{(}\hskip0pt\crcr\noalign{\kern3pt}}}}}
\newcommand{\cbp}{\mathop{\vtop{ialign{##\crcr
  $\hfil\displaystyle{}\hfil$\crcr\noalign{\kern-13pt\nointerlineskip}
  \big{)}\hskip0pt\crcr\noalign{\kern3pt}}}}}

\newcommand{\s}{\sigma}
\newcommand{\srange}{\sigma=0,...,N_c-1}

\renewcommand{\sp}{,\hspace{.3in}}

\newcommand{\p}{^\prime}

\newcommand{\w}{\omega}

\newcommand{\hc}{$\hat{J}_{\gst}$}

\newcommand{\ad}{\frac{A(\dl)}{\al}}

\newcommand{\sgb}{{\mbox{\scriptsize{\gb}}}}

\def\gb            {\mbox{$\hat{\mathfrak g}$}}
\def\hb            {\mbox{$\hat{\mathfrak h}$}}

\def\sm#1      {\mbox{\scriptsize $#1$}}

\def\md        {\mbox{\Large $\mathbb  D$}}
\def\srac#1#2{\smal{\frac{#1}{#2}}}

\def\smal#1{\mbox{\small $#1$}}
\def\big#1{\mbox{\large $#1$}}
\def\Big#1{\mbox{\Large $#1$}}

\def\hjb{\hat{\bar{J}}}

\newcommand{\nb}{\bar n(r)}
\def\hjbb{ \hat{\bar{J}}^\sharp }

\def\hc{^\dagger}

\def\one{{\mathchoice {\rm 1\mskip-4mu l} {\rm 1\mskip-4mu} {\rm 1\mskip-4.5mu l} {\rm 1\mskip-5mu l}}}

\def\mmrrs{m+\srac{n(r)}{\r(\s)}}
\def\nnsrs{n+\srac{n(s)}{\r(\s)}}

\def\nrm{{n(r)\m}}

\def\nrn{{n(r)\n}}
\def\mnrn{{-n(r),\n}}
\def\nsn{{n(s)\n}}

\def\mnnrnsrsf{{m+n+\frac{n(r)+n(s)}{\r(\s)}}}

\def\sG{{\cal G}}

\def\hj{\hat{J}}

\def\nrn{{n(r)\n}}
\def\nsn{{n(s)\n}}

\def\hc{^\dagger}
\def\st{{\cal T}}
\def\0b{\ }
\def\pl{\partial}

\def\srange{\s=0,\ldots,N_c-1}
\def\sm{{\cal M}}

\def\hb{{\hat \beta}}
\def\skl{\vskip .3cm \ni}
\def\tone{{ \rm 1\mskip-4.5mu l}}

\def\irrep#1{{\bf \mathfrak{#1}}}
\def\thickone{{\rm 1\mskip-4.5mu l}}
\newcounter{myfigctr}
\def\myfig#1{\refstepcounter{myfigctr}%
 \label{#1}%
} 

\makeatletter
\renewcommand{\@makefnmark}{\mbox{$^{\ddagger\@thefnmark}$}}
\renewcommand{\subsection}{\@startsection
  {subsection}{2}{0pt%-1em
}{-\baselineskip}{0.5\baselineskip}
   {\normalfont\normalsize\bf}}
\renewcommand{\section}{\@startsection
  {section}{2}{0pt%-1em
}{-\baselineskip}{0.5\baselineskip}
  {\bf\large}}

\makeatother
\numberwithin{equation}{section}
\newcommand{\publititle}[8]
{ %\clearemptydoublepage
  \vspace*{-3cm}
  \begin{flushright} #1 \\ {\tt #2} \end{flushright}
  \vfill
  \begin{center}{\Large%\scshape
    \bfseries #3}\end{center}
  \vskip 8mm
  \begin{center}{\large #4}\end{center}
  \begin{center}{\normalsize #5}\end{center}
  \vskip 8mm
  \nopagebreak
  \noindent #6
  \vfill
  \begin{flushleft} #7
  \end{flushleft}
  \hrule width 6.7cm \vskip.1mm
  {\small #8}
  \thispagestyle{empty}
  \clearpage
}

\begin{document}

\begin{titlepage}

\rightline{\vbox{\small\hbox{UCB-PTH-02/43} \hbox{LBNL-51609}
 \vskip.5ex  \hbox{\tt hep-th/0211003} \hbox{\today} }}
\vskip 2.0cm

\centerline{\LARGE \bf The Outer-Automorphic  WZW Orbifolds on $\so (2n)$,}
\centerline{\LARGE \bf
including Five Triality Orbifolds on $\so (8)$} \vskip 1.5cm
 \centerline{{\Large   O. Ganor${}^a$,
M.~B. Halpern${}^a$, C. Helfgott${}^a$, N.~A. Obers${}^b$ } }
%\centerline{{\bf People}}
\vskip 1cm
\begin{center}
{\sl ${}^a$Department of Physics, University of California, Berkeley,
California 94720, USA
 \\[.3ex] and  \\[.3ex] Theoretical Physics Group,  Lawrence Berkeley National
Laboratory \\
     University of California,
     Berkeley, California 94720, USA \\[.75ex] $^b$The
Niels Bohr Institute, Blegdamsvej 17, DK-2100 Copenhagen \O, Denmark}
\vskip .1in  {\small \sffamily origa@socrates.berkeley.edu,
halpern@physics.berkeley.edu, \\
helfgott@socrates.berkeley.edu, obers@nbi.dk }
\end{center}

\vskip 0.8cm  \centerline{\bf \large Abstract} \vskip 0.1cm
\noindent Following recent advances in the local theory of
current-algebraic orbifolds we present the basic dynamics
- including the {\it twisted KZ equations} - of each twisted sector
of all outer-automorphic  WZW orbifolds on
$\so(2n)$. Physics-friendly Cartesian bases are used throughout,
and we are able in particular to assemble two $\Zm_3$ triality
orbifolds and three $S_3$ triality orbifolds on $\so(8)$.

\end{titlepage}

 \clearpage

\renewcommand{\baselinestretch}{.4}\rm
{\footnotesize
\tableofcontents
}

\renewcommand{\baselinestretch}{1.2}\rm

\section{Introduction}

In the last few years there has been a quiet revolution
 in the local theory of {\it current-algebraic orbifolds}.
Building on the discovery of {\it orbifold affine algebras}
\cite{Borisov:1997nc,Evslin:1999qb} in the cyclic
permutation orbifolds, Refs.~\cite{deBoer:1999na,Evslin:1999ve,Halpern:2000vj}
 gave the twisted currents and stress tensor in
 each twisted sector of any current-algebraic
orbifold $A(H)/H$ - where $A(H)$ is any current-algebraic conformal
field theory [6-11]
%\cite{Bardakci:1971nb,Halpern:1971ay,Halpern:1971qj,Halpern:1989ss,Halpern:1992gb,Halpern:1996js}
 with a finite symmetry group $H$. The
construction treats all current-algebraic orbifolds at the same time, using the
method of {\it eigenfields} and the
{\it principle of local isomorphisms} to map OPEs in the symmetric theory to OPEs
 in the orbifold. The orbifold results are
expressed in terms of a set of twisted tensors or {\it duality transformations},
 which are discrete Fourier transforms constructed
from the eigendata of the $H$-{\it eigenvalue problem}.

More recently, the special case of the WZW orbifolds $A_g(H)/H ,\, 
H\subset Aut(g)$ was worked out in further detail
\cite{deBoer:2001nw,Halpern:2002ww,Halpern:2002vh}, introducing
the {\it extended $H$-eigenvalue problem} and the
{\it linkage relation} to include the {\it twisted affine
primary fields} and the {\it twisted KZ equations} of the
WZW orbifolds. The general form of the twisted KZ equations
is reviewed in Sec.~\ref{seckz}. There has also been progress in WZW orbifolds
at the level of characters \cite{KacTod,Borisov:1997nc,Ban,Birke}.

In addition to the operator formulation, the {\it general WZW orbifold action}
was also given in Ref.~\cite{deBoer:2001nw},
with applications to special cases in Refs.~\cite{deBoer:2001nw,Halpern:2002ww,Halpern:2002vh}.
 The general WZW orbifold action provides
the classical description of each sector of every WZW orbifold in terms of
appropriate {\it group orbifold elements},
which are the classical limit of the twisted affine primary fields. Moreover,
Ref.~\cite{Halpern:2002zv} gauged the
general WZW orbifold action by general twisted gauge groups to obtain the
{\it general coset orbifold action}.

In this paper we illustrate the general formulation of WZW orbifolds
by working out another large class of examples in further detail.
In particular, we use physics-friendly Cartesian bases to
describe the basic dynamics, including the twisted affine Lie
algebras, the twisted affine-Sugawara constructions and the twisted
KZ equations of all the outer-automorphic
WZW orbifolds on $\so(2n) \cong \mathfrak{spin}(2n)$. This includes two $\Zm_3$ triality
orbifolds and, somewhat surprisingly, three $S_3$ triality orbifolds
on $\so (8)\cong \mathfrak{spin}(8)$.

Some remarks about general WZW orbifolds are also included.
In particular, Subsec.~\ref{recsec} completes the solution of the
rectification problem \cite{deBoer:2001nw} for all the basic
twisted right-mover current algebras, and Subsec.~\ref{twassec} completes
the computation of the scalar twist-field conformal weights
for all sectors of all the basic WZW orbifolds.

\section{The Twisted KZ Systems of the WZW Orbifolds \label{seckz}}

We begin by reminding the reader that twisted KZ systems are now
known \cite{deBoer:2001nw,Halpern:2002ww,Halpern:2002vh} for the
correlators in the {\it scalar twist-field state}%
 \footnote{For the inner-automorphic WZW orbifolds a different
 set of twisted KZ equations \cite{deBoer:2001nw} was given for
 the correlators in the untwisted affine vacuum state.}
  $|0\rangle_\s = \tau_\s(0) | 0\rangle$
\begin{subequations}
\begin{equation}
\hat A_+ (\T,z,\s) \equiv {}_\s\langle 0|  \hgp(\T^{(1)},z_1,\s) \hgp(\T^{(2)},z_2,\s)
\cdots \hgp(\T^{(N)},z_N,\s) |0 \rangle_\s
\end{equation}
\begin{equation}
\hat A_- (\T,\bz,\s) \equiv {}_\s\langle 0|  \hgm(\T^{(1)},\bz_1,\s) \hgm(\T^{(2)},\bz_2,\s)
\cdots \hgm(\T^{(N)},\bz_N,\s) |0 \rangle_\s
\end{equation}
\begin{equation}
\label{vacl}
\hj_{\nrm } ( m + \srac{n(r)}{\r (\s)} \geq 0) | 0 \rangle_\s
= {}_\s \langle 0 | \hj_{\nrm } ( m + \srac{n(r)}{\r (\s)} \leq 0)
=0
\end{equation}
\begin{equation}
\label{vacr}
\hjb_{\nrm } ( m + \srac{n(r)}{\r (\s)} \leq 0) | 0 \rangle_\s
= {}_\s \langle 0 | \hjb_{\nrm } ( m + \srac{n(r)}{\r (\s)} \geq 0)
=0
\end{equation}
\begin{equation}
\srange
\end{equation}
\end{subequations}
in each sector $\s$ of any WZW orbifold $A_g(H)/H$. Here $N_c$ is
the number of conjugacy classes of the original symmetry group $H$,
$\hj_\nrm (m+\srac{n(r)}{\r(\s)})$ are the modes of the twisted
current algebra \cite{Halpern:2000vj,deBoer:2001nw} of that
sector and
\begin{equation}
\hg  (\T,\bz,z,\s) = \hgm (\T,\bz,\s) \hgp (\T,z,\s)
\end{equation}
is the {\it twisted affine primary field} in twisted representation
$\T = \T (T,\s)$ of sector $\s$. The sector $\s=0$ is the symmetric
theory $A_g(H)$, $H \subset {\rm Aut}\,g$ where the currents and
affine primary fields are untwisted and the scalar twist-field
state $| 0 \rangle_0 = | 0 \rangle$ is the untwisted affine vacuum.

The scalar twist-field states exist for all sectors of all
current-algebraic orbifolds.
The reason \cite{Halpern:2002vh} is easy to understand in the equivalent form
\begin{equation}
\hj_{\nrm } ( m + \srac{n(r)}{\r (\s)} >0) | 0 \rangle_\s = 0
\sp \hj_{0\m} (0)| 0 \rangle_\s  =0 \ .
\end{equation}
The first condition holds for any primary state of any
(infinite-dimensional) Lie algebra and the second condition
restricts our attention to the ``s-wave'' or trivial  representation
of the residual (untwisted) symmetry of the sector.
For the WZW permutation orbifolds and the outer-automorphic
WZW orbifolds on simple $g$, the scalar twist-field state is
also known to be the ground state of each sector.

For each sector $\s$ of any WZW orbifold $A_g(H)/H$,
the twisted left-and right-mover KZ systems are
\begin{subequations}
\label{tlmkzeq}
\begin{equation}
\part_\kappa \hat A_+(\T,z,\s) = \hat A_+ (\T,z,\s) \hat W_\kappa (\T,z,\s) \sp
\sp \kappa = 1 \ldots N \sp \s = 0, \ldots ,N_c -1
\end{equation}
\begin{equation}
\label{kzctl}
\hat W_{\kappa}(\T,z,\s) = 2 \lr^{n(r)\m;-n(r),\n} (\s)
 \left[ \sum_{\r \neq \k}\left( \frac{z_\r}{z_\k}
\right)^{ \srac{\bar n(r)}{\r(\s)}} \frac{1}{z_{\k \r }}
\T_{n(r)\m}^{(\r)} \T_{-n(r),\n}^{(\k)}
- \srac{\bar n(r)}{\r(\s)} \frac{1}{z_{\k}}
\T_{n(r)\m}^{(\k)} \T_{-n(r),\n}^{(\k)} \right]
\end{equation}
\end{subequations}
\begin{subequations}
\label{trmkzeq}
\begin{equation}
\bar \part_\kappa \hat A_-(\T,\bz,\s) =
\hat{\bar{W}}_\kappa (\T,\bz,\s) \hat A_- (\T,\bz,\s) \sp
\sp \kappa = 1 \ldots N \sp \s = 0, \ldots ,N_c -1
\end{equation}
\begin{equation}
\label{kzctr}
\hat{\bar{W}}_{\kappa}(\T,\bz,\s)
= 2 \lr^{n(r)\m;-n(r),\n} (\s)
 \left[ \sum_{\r \neq \k}\left( \frac{\bz_\r}{\bz_\k}
\right)^{ \srac{\nb}{\r(\s)}} \frac{1}{\bz_{\k \r }}
\T_{n(r)\m}^{(\k)} \T_{-n(r),\n}^{(\r)}
- \srac{\nb}{\r(\s)} \frac{1}{\bz_{\k}}
\T_{n(r)\m}^{(\k)} \T_{-n(r),\n}^{(\k)} \right]
\end{equation}
\end{subequations}
\begin{subequations}
\label{gwig0}
\begin{equation}
\label{gwig}
\hat A_+ (\T,z,\s) \left(\sum_{\r =1}^N \T_{0,\m}^{(\r)}\right) =
\left(\sum_{\r =1}^N \T_{0,\m}^{(\r)}\right)  \hat A_- (\T,\bz,\s) = 0 \sp \forall \;\m
\end{equation}
\begin{equation}
\T^{(\r)} \T^{(\k)} \equiv \T^{(\r)}\otimes \T^{(\k)} \sp
z_{\k \r} \equiv z_\k - z_\r \sp \bz_{\k \r} \equiv \bz_\k - \bz_\r
\ .
\end{equation}
\end{subequations}
Here $\r (\s)$ is the order of the automorphism $h_\s \in H$
and the integers $n(r)$ and their pullbacks $\bar{n}(r)$
are obtained from the {\it $H$-eigenvalue problem} of sector $\s$.
The twisted tensors $\lr (\s)$ and
$\T = \T(T,\s)$ are respectively the {\it twisted inverse inertia
tensor} and the {\it twisted representation matrices}, formulas
for which are given in Ref.~\cite{deBoer:2001nw}, and the relations
in \eqref{gwig} are the Ward identities of the residual symmetry
algebra of each sector $\s$.
In the untwisted sector $\s=0$, the twisted KZ systems reduce
to the ordinary KZ system \cite{Knizhnik:1984nr} of the symmetric theory $A_g(H)$.

The explicit form of the general twisted KZ system
\eqref{tlmkzeq}-\eqref{gwig0} has so far
 been worked out for the following cases: \nl
$\bullet$ the WZW permutation orbifolds
\cite{deBoer:2001nw,Halpern:2002ww,Halpern:2002vh} \nl
$\bullet$ the inner-automorphic WZW orbifolds
\cite{Halpern:2002vh} on simple $g$ \nl
$\bullet$ the (outer-automorphic) charge conjugation
orbifold on $\su (n\geq 3)$ \cite{Halpern:2002ww}. \nl
Ref.~\cite{Halpern:2002ww} also solved the twisted vertex
operator equations and the twisted KZ systems in an abelian
limit to obtain the twisted vertex operators for each sector
of a very large class of abelian orbifolds%
\footnote{An abelian twisted KZ equation for the inversion
orbifold  $x \rightarrow -x$ was given  earlier in Ref.~\cite{Frohlich:1999ss}.}.
The WZW permutation orbifolds also exhibit an extended operator algebra
including {\it twisted Virasoro generators}
\cite{Borisov:1997nc,Dijkgraaf:1997vv,Halpern:2002vh} and
finally - with
emphasis on the WZW permutation orbifolds - the general
twisted KZ system has been used to study the reducibility
\cite{Halpern:2002vh} of the general twisted affine primary field.

In what follows, we apply the general theory of WZW orbifolds to
work out the basic dynamics of all the
outer-automorphic WZW orbifolds on $\so(2n)$, including the triality
orbifolds on $\so(8)$. The explicit forms of all the relevant
twisted KZ systems are found in Subsec.~\ref{kzsec}. Except for our
discussion of the rectification problem in Subsec.~\ref{recsec}
and the action formulation in Subsec.~\ref{secelt},
we concentrate for brevity on the twisted left-mover systems -
but the twisted right-mover systems
\cite{deBoer:2001nw,Halpern:2002ww,Halpern:2002vh}
can be easily evaluated with the same data.

\section{The Currents and Symmetries of $\so (2n)$ and $\so (8)$
\label{symsec}}

For accessibility in physics we begin in the standard Cartesian
basis of $\so (2n)$, where the OPEs of the
currents of affine $\so (2n)$ are
\begin{subequations}
\label{cartbas}
\begin{equation}
\label{sonca}
J_{ij} (z) J_{kl} (w) =\frac{2 k (e_{ij})_{kl}}{(z-w)^2}
+ \frac{i}{z-w} ( \de_{jk} J_{il} (w) + \de_{il} J_{jk} (w)
-\de_{ik} J_{jl} (w) - \de_{jl} J_{ik} (w)) + \Ozw
\end{equation}
\begin{equation}
\label{eijdef}
(e_{ij})_{kl} \equiv \frac{1}{2} (\de_{ik} \de_{jl} - \de_{jk} \de_{il})
\sp i,j = 1 \ldots n \sp J_{ij} (z) = -J_{ji} (z) \ .
\end{equation}
\end{subequations}
Here we have chosen root length $\psi^2=2$ for $\so(2n \geq 4)$
and the invariant
level of $\so (2n \geq 4)$ is $x = 2k /\psi^2 = k$. The antisymmetric
tensors $\{ e_{ij} \}$ satisfy
\begin{equation}
(e_{ij})_{kl} J_{kl}(z) = J_{ij}(z) \sp (e_{ij})_{ij} = \frac{1}{2}
\sp
{\rm Tr}(e_{ij} e_{kl}) = - (e_{ij})_{kl}
\end{equation}
and the $n(2n-1)$ independent currents of $\so (2n)$ are obtained by
choosing $1 \leq i < j \leq 2n$. In what follows we discuss
the action of the various outer automorphisms (see the table in
App.~\ref{appinvsub}) of $g=\so(2n)$ and $\so(8)$ in the notation of
Refs. \cite{deBoer:1999na,Halpern:2000vj}
\begin{equation}
J_{ij} (z)' = \sum_{k,l} \o_{ij,kl} J_{kl} (z) \equiv
 \o_{ij,kl} J_{kl} (z) \sp \o_{ij,kl} = - \o_{ji,kl} = - \o_{ij,lk}
\end{equation}
where the matrix $\o \equiv \o (h_\s)$ is the action of the automorphism
$h_\s \in H \subset {\rm Aut}\,g$ in the adjoint.
Moreover, because it will facilitate transition to the orbifold, we
will insist in this paper on finding bases in which the action $\o$
of each automorphism is {\it diagonal}.

\subsection{The parity automorphism $\Pm$ \label{parsec}}

As defined below, the parity automorphism $\Pm $  on $\so (2n)$ is
a $\Zm_2$-type
outer automorphism for all $2n \geq 6$. For $ 2n \geq 8$, $\Pm$
is equivalent to the exchange of the two rightmost nodes of the
$D_n$ Dynkin diagrams and for $\so (6)$ it is equivalent to the
exchange of the outer nodes of the Dynkin diagram of $\su(4)$.
Then we know that, up to a unitary transformation,
$\Pm$ maps $S \leftrightarrow C$ where $S$ and $C$
\begin{subequations}
\begin{equation}
\mu_S = \frac{1}{2} \sum_{i=1}^n \s_i e_i \sp \s_i = \pm
\sp \mbox{(even number of +)}
\end{equation}
\begin{equation}
\mu_C = \frac{1}{2} \sum_{i=1}^n \s_i e_i \sp  \s_i = \pm
\sp \mbox{(odd number of +)}
\end{equation}
\end{subequations}
are the Weyl spinor reps of $\mathfrak{spin}(2n)$ with opposite
chirality (see also App.~\ref{apprep}).

To obtain  a diagonal basis for $\Pm$, we label the currents as follows
\begin{equation}
J_{ij} (z) = \{ J_{\m \n} (z), J_\m (z) \equiv J_{\m ,2n} (z) \}
\sp \m ,\n = 1 , \ldots , 2n -1
\end{equation}
where the set $\{ J_{\m \n}(z) \}$ generates an affine $\so (2n-1)$ at the
same invariant level $x$. In this basis, the OPEs of affine
$\so (2n)$ read
\begin{subequations}
\label{spinrep}
\begin{equation}
J_{\m \n} (z) J_{\r \s} (w) =\frac{2 k (e_{\m \n})_{\r \s}}{(z-w)^2}
+ \frac{i}{z-w} ( \de_{\n \r} J_{\m \s} (w) + \de_{\m \s} J_{\n \r} (w)
-\de_{\m \r } J_{\n \s} (w) - \de_{\n \s} J_{\m \r} (w)) + \Ozw
\end{equation}
\begin{equation}
J_{\m \n} (z) J_{\r } (w) =
 \frac{i}{z-w} ( \de_{\n \r} J_{\m } (w) -\de_{\m \r} J_{\n } (w) )
+ \Ozw
\end{equation}
\begin{equation}
J_{\m } (z) J_{\n } (w) = \frac{k \de_{\m \n}}{(z-w)^2}
 -\frac{i}{z-w} J_{\m \n } (w) + \Ozw \ .
\end{equation}
\end{subequations}
The automorphism $\Pm $ acts on the currents as
\begin{subequations}
\label{omP0}
\begin{equation}
\label{omP}
\o_{\m \n,\r \s} = (e_{\m \n} )_{\r \s} \sp
\o_{\m,\n} = - (e_{\m,2n})_{\n,2n} = - \de_\m^\n
\end{equation}
\begin{equation}
 J_{\m \n} (z)' = J_{\m \n} (z) \sp
J_\m (z)' = -J_\m (z)
\end{equation}
\end{subequations}
where the sign reversal of the vector current is a ``parity''
transformation.
This action corresponds to the coset decomposition
$g  = h + g/h$, where $h$ is the invariant subalgebra and
\begin{equation}
\label{Pmss}
\frac{g}{h} = \frac{\so (2n)_x}{\so (2n -1)_x} \sp n \geq 3
\end{equation}
is a symmetric space. As seen in the table of App.~\ref{appinvsub},
these symmetric spaces
correspond to the outer automorphisms in entry iv) and the
``alternate'' charge conjugation automorphism
(see Ref.~\cite{Halpern:2002ww}) on $\su(4)$
\begin{equation}
\frac{\so(6)}{\so(5)} = \frac{\su(4)}{C_2}
= \frac{\su(4)}{\mathfrak{sp}(2)}
\end{equation}
given in entry iii) of the table.
Note that all $\Zm_2$-type automorphisms, inner or outer,
correspond to symmetric space decompositions, and vice-versa,
but not all
symmetric space decompositions correspond to outer automorphisms.

\subsection{The charge-conjugation automorphism $\Cm$}

For any Lie $g$, the charge-conjugation automorphism $\Cm \in
\Zm_2$ is equivalent  to sign reversal $\{ \m (T) \rightarrow
- \m (T) \}$ of all weights $\m$ of any irrep $T$. Correspondingly,
the action $T'$ of $\Cm$ on any matrix irrep $T$ of Lie $g$ is
unitarily equivalent to $\bar T$
\begin{equation}
T' \equiv \o T \cong \bar T \equiv - T^{\rm t}
\end{equation}
where ${\rm t}$ is matrix transpose. See Ref.~\cite{Halpern:2002ww}
for the diagonal action of  $\Cm$ in the standard Cartesian basis of $\su(n)$.

To obtain a diagonal basis for $\Cm$ on  $\so( 2n)_x$, we label
the currents as follows
\begin{subequations}
\begin{equation}
J_{ij}(z) = \{ J_{AB} (z) , J_{IJ} (z) , J_{AI} (z) \} 
\end{equation}
\begin{equation}
i = (A,I) \sp A = 1 \ldots n \sp I = n+1 , \ldots , 2n
\end{equation}
\end{subequations}
so that the zero modes of the currents $\{J_{AB} (z),J_{IJ} (z)\}$ generate
the subalgebra $(\so(n) \oplus \so(n)) \subset \so(2n)$.
In this basis, the OPEs  of affine $\so (2n)$  read
\begin{subequations}
\label{CmOPE}
\begin{eqnarray}
J_{AB} (z) J_{CD} (w)\hskip -.2cm &=&\hskip -.1cm\frac{2 k (e_{AB})_{CD}}{(z-w)^2}
+ \frac{i}{z-w} (\de_{BC} J_{AD}(w) + \de_{AD} J_{BC} (w)
 -\de_{AC} J_{BD} (w)- \de_{BD} J_{AC}(w) ) \nn \\
  & & \hskip 1cm + \Ozw
\end{eqnarray}
\begin{eqnarray}
J_{IJ} (z) J_{KL} (w) \hskip -.2cm& = & \hskip -.1cm \frac{2 k (e_{IJ})_{KL}}{(z-w)^2}
+ \frac{i}{z-w} (\de_{JK} J_{IL}(w) + \de_{IL} J_{JK} (w) \
 -\de_{IK} J_{JL} (w)- \de_{JL} J_{IK}(w) ) \nn \\
 && \hskip 1cm + \Ozw
\end{eqnarray}
\begin{equation}
J_{AB}(z)J_{IJ} (w) = \Ozw
\end{equation}
\begin{equation}
J_{AB}(z ) J_{CI} (w)
 =   \frac{i}{z-w} (\de_{BC} J_{AI}(w) - \de_{AC} J_{BI} (w))+\Ozw
\end{equation}
\begin{equation}
J_{IJ}(z) J_{AK} (w)
 =   \frac{i}{z-w} (\de_{JK} J_{AI}(w) - \de_{IK} J_{AJ} (w))+\Ozw
\end{equation}
\begin{equation}
J_{AI}(z) J_{ BJ} (w)
= \frac{k \de_{AB} \de_{IJ} }{(z-w)^2}
 - \frac{i}{z-w} (\de_{AB} J_{IJ}(w) + \de_{IJ} J_{AB} (w))+\Ozw \ .
\end{equation}
\end{subequations}
The diagonal action of  $\Cm $ in this basis
\begin{subequations}
\label{Cact}
\begin{equation}
\label{omC}
\o_{AB,CD} = (e_{AB})_{CD} % \de_A^C \de_B^D
\sp \o_{IJ,KL} =(e_{IJ})_{KL} %  \de_I^K \de_J^L
\sp \o_{AI,BJ} = -(e_{AI})_{BJ} = - \de_{AB} \de_{IJ}
\end{equation}
\begin{equation}
\label{chargeaut}
 J_{AB} (z)' =J_{AB} (z) \sp  J_{IJ} (z)' =J_{IJ} (z) \sp
 J_{AI} (z)' = -J_{AI} (z)
 \end{equation}
\end{subequations}
is an automorphism of the OPEs \eqref{CmOPE}. The action \eqref{Cact}
defines the symmetric space
\begin{equation}
\label{Cmss}
\frac{g}{h} = \frac{\so (2n)_x}{\so (n)_{x\tau(n)} \oplus
\so(n)_{x\tau (n)}} \sp
\tau (n) = \left\{ \begin{array}{ll} 2 & \mbox{for}\,\; n =3
\\ 1  & \mbox{for}\,\; n \geq 4 \end{array} \right.
\end{equation}
with invariant subalgebra $h= \so(n) \oplus \so (n)$.

To see that $\Cm$ is equivalent to charge conjugation on $\so (2n)
\cong \spi (2n)$,
note that the choice of Cartan subalgebra
$\{ J_{i,2n -i} , \ i = 1 \ldots n \}$ is entirely in $g/h$.
Then the action of $\Cm$ on these Cartan currents is
\begin{equation}
\label{CartanC}
J_{i,2n-i}(z)' = - J_{i,2n-i}(z) \sp i = 1 \ldots n
\end{equation}
and, as required by charge conjugation, $\Cm$ reverses the sign
of all weights as measured by this choice of Cartan subalgebra.

Of course, this choice is equivalent to any other choice of Cartan
subalgebra, for example the standard choice
$\{ J_{2i -1,2i} , \ i = 1 \ldots n \}$.
 For $\so (4r)$ the action of $\Cm$ on these Cartan currents is
\begin{equation}
\label{CartanC2}
J_{2i-1,2i}(z)' = J_{2i-1,2i}(z) \sp i = 1 \ldots n \sp n = 2 r
\end{equation}
so that sign reversal of all weights is equivalent to no sign
reversal and
\begin{equation}
T ' \cong \bar T \cong T \sp \forall \;\, T \;\,\mbox{on}\;\, \so(4r)
\cong \spi (4r) \ .
\end{equation}
This is in agreement with the table in App.~\ref{appinvsub},
which tells us that
\begin{subequations}
\label{inout}
\begin{eqnarray}
\bullet  & & \Cm = \mbox{inner automorphism of } \so (4r)
\hskip 2cm
\\
\bullet  & & \Cm \simeq \Pm = \mbox{outer automorphism of }
\so (4r+2) \label{inoutb}
\end{eqnarray}
\end{subequations}
where $\Pm$ is the Dynkin automorphism of $\so (2n)$. More precisely,
\eqref{inoutb} means that $\o (\Cm) = K \o (\Pm)$ where $K$ is an inner
automorphism which acts non-trivially.
Another way to understand \eqref{inout} is as follows:
Since $\Cm$ reverses the sign of each weight, it is easy to
check from the weights \eqref{spinrep} that $\Cm$ maps
\begin{subequations}
\begin{equation}
S \leftrightarrow C \qquad \mbox{for } \so (4r+2) \cong \spi (4r+2)
\end{equation}
\begin{equation}
S \rightarrow S \sp C \rightarrow C  \qquad \mbox{for } \so (4r)
\cong \spi (4r)
\end{equation}
\end{subequations}
which confirms the statement in \eqref{inout}.

\subsection{The interpolating automorphisms $\Cmg$ \label{intsec}}

We consider next a set of $\Zm_2$-type automorphisms of $\so(2n)$
\begin{equation}
\label{interaut}
\Cmg \sp r = n,\ldots,2n-1 \ : \qquad
\frac{g}{h} = \frac{\so (2n)_x}{\so (r)_{ x\tau (r)} \oplus
\so (2n-r)_{x\tau (2n-r)}}
\end{equation}
which, as we shall see, interpolates between the automorphisms
$\Pm$ and $\Cm$ described above. More precisely, we find that
the ``endpoints'' of the sequence are
\begin{equation}
\mathbb{A} (2n;n) = \Cm \sp\mathbb{A} (2n;2n-1) = \Pm
\end{equation}
as one might expect by comparison of the symmetric spaces in
\eqref{interaut} with those in \eqref{Pmss} and \eqref{Cmss}.

To describe the action of $\Cmg$, it is convenient to employ the
same notation used above for $\Cm$, now with the more general
ranges
\begin{subequations}
\label{omCg0}
\begin{equation}
\Cmg \sp r = n, \ldots,2n-1
\end{equation}
\begin{equation}
J_{ij}(z) = \{ J_{AB} (z) , J_{IJ} (z) , J_{AI} (z) \} \sp
\end{equation}
\begin{equation}
i = (A,I) \sp A = 1 \ldots  r \sp I = r+1 , \ldots , 2n
\end{equation}
\begin{equation}
\label{omCg}
\o_{AB,CD} = (e_{AB})_{CD} % \de_A^C \de_B^D
\sp \o_{IJ,KL} =(e_{IJ})_{KL} %  \de_I^K \de_J^L
\sp \o_{AI,BJ} = -(e_{AI})_{BJ} = - \de_{AB} \de_{IJ}
\end{equation}
\begin{equation}
\label{chargeautg}
 J_{AB} (z)' =J_{AB} (z) \sp  J_{IJ} (z)' =J_{IJ} (z) \sp
 J_{AI} (z)' = -J_{AI} (z) \ .
 \end{equation}
\end{subequations}
Bearing these ranges in mind,  the OPEs \eqref{CmOPE} are in
a diagonal basis for all $\Cmg$.

According to  the  table of App.~\ref{appinvsub}, the
automorphism $\mathbb{A}(2n;r)$ is an outer automorphism of $\so (2n)$
only when $r$ is odd, and an inner automorphism when $r$
is even. This gives for example the $\Zm_2$-type outer automorphisms
\begin{subequations}
\begin{equation}
\Am (6;3) \quad : \quad
\frac{\so(6)_x}{\so(3)_{2x} \oplus \so(3)_{2x}} \sp
\Pm = \Am (6;5) \quad : \quad
\frac{\so(6)_x}{\so(5)_{x} }
\end{equation}
\begin{equation}
\Am (8;5) \quad : \quad
\frac{\so(8)_x}{\so(5)_{x} \oplus \so(3)_{2x}} \sp
\Pm = \Am (8;7) \quad : \quad
\frac{\so(8)_x}{\so(7)_{x}}
\end{equation}
\begin{equation}
\Am (10;5) \;\; : \; \;\;
\frac{\so(10)_x}{\so(5)_{x} \oplus \so(5)_{x}} \sp
\Am (10;7)\;\; : \; \;\;
\frac{\so(10)_x}{\so(7)_{x} \oplus \so(3)_{2x} } \sp
 \Pm = \Am (10;9)\;\; : \; \;\;
\frac{\so(10)_x}{\so(9)_{x}}
\end{equation}
\end{subequations}
for $\so(6)$, $\so(8)$ and $\so(10)$. More generally,
$\{ \Am (2n;r) \}$ contains $m$ distinct $\Zm_2$-type outer
automorphisms for $\so (4m)$ and $m+1$ for $\so (4m+2)$, and this
list includes {\it all} the $\Zm_2$-type outer automorphisms given
in the table for $\so (2n)$.

Since all the $\Zm_2$-type outer automorphisms are inner-automorphically
equivalent to the $\Zm_2$-Dynkin automorphism, we have
\begin{subequations}
\begin{equation}
\Cm = \mathbb{A} (2n;n) \simeq\mathbb{A} (2n;n+2) \simeq
\cdots \simeq \mathbb{A} (2n;2n-1) = \Pm \sp \mbox{for}\;\, n\;\,
\mbox{odd}
\end{equation}
\begin{equation}
 \mathbb{A} (2n;n+1) \simeq\mathbb{A} (2n;n+3) \simeq
\cdots \simeq \mathbb{A} (2n;2n-1) = \Pm \sp \mbox{for}\;\, n\;\,
\mbox{even} \ .
\end{equation}
\end{subequations}
We emphasize however that each automorphism $\Am (2n,r)$,
$r= n,\ldots,2n-1$, inner or outer, gives rise to a physically
distinct twisted sector.

We also note that the parity automorphism $\Pm =\mathbb{A} (2n;2n-1)$
is the only automorphism in $\Cmg$ which is an outer
automorphism of $\so (2n)$ for all $n$. Moreover,  the
charge-conjugation automorphism $\Cm = \mathbb{A} (2n;n)$ is the only
automorphism in $\Cmg$ for which a full set of Cartan  generators
(see Eq.~\eqref{CartanC}) is in $g/h$ - and hence the only one which is
equivalent to sign reversal of all the weights of each $\spi (2n)$.

In what follows, we defer to the mathematicians by treating
the case of $\Pm =\mathbb{A} (2n;2n-1)$ separately, but we will
often include $\Cm = \mathbb{A} (2n;n)$ as a special case of
the full set $\Cmg$.

\subsection{The triality automorphism $\Tm$ \label{T1sec}}

In this subsection we consider the first triality automorphism
$\Tm \in \Zm_3$ which, as we shall see, is equivalent to the
Dynkin automorphism in entry viii) of the table in
App.~\ref{appinvsub}.
In this discussion we will use the notation
\begin{equation}
\{ J_{ij} (z) \} \sp 1 \leq i < j \leq 8 \quad ; \quad
\{ J_{\mu \nu } (z) \} \sp 1 \leq \m < \n  \leq 7
\end{equation}
for the currents of $\so (8)$ and $\so (7)$ respectively.

We recall first the sequence of maximal subalgebras
\begin{equation}
\so (8)_x \supset \so (7)_x \supset (\gt)_x
\end{equation}
each with embedding index one. Refs.~\cite{Humphreys:1972,Dundarev:1984}
include discussions of this embedding, but we can work out what we need
from the fact (see e.g. Ref.~\cite{Georgi:1982})
that $\gt$ is the subalgebra of $ \so (7)$ which
leaves invariant the octonionic structure constants.

For the octonions $\{1, i_\m \}$ we will use the basis
\begin{subequations}
\begin{equation}
i_\m i_\n = - \de_{\m \n} + g_{\m \n \r} i_\r \sp
\m , \n = 1 \ldots 7
\end{equation}
\begin{equation}
\label{oct}
g_{123}= g_{247} = g_{451} = g_{562} = g_{634} = g_{375} = g_{716} = 1
\sp g_{\a \r \s} g_{\be \r \s} = 6 \de_{\a \be}
\end{equation}
\end{subequations}
and then we need to solve for the 14-dimensional subspace
of $\gt$ currents $\{ J_A(z) \}$
\begin{subequations}
\begin{equation}
\label{Jgt}
J_A (z)= (\r_A)_{\m \n} J_{\m \n} (z)\sp A= 1 \ldots 14
\end{equation}
\begin{equation}
\label{g2inv}
(\r_A)_{ \a \a'} g_{\a'\be \ga} +
(\r_A)_{ \be \be '} g_{\a\be' \ga} +
(\r_A)_{ \ga \ga'} g_{\a\be \ga'} = 0 \sp
(\r_A)_{\n \m} = -(\r_A)_{\m \n} \ .
\end{equation}
\end{subequations}
An explicit form of the fourteen $\r_A$'s in a
trace-orthogonal basis
\begin{equation}
\label{rhogrel}
{\rm Tr} (\r_A \r_B) = -\frac{1}{2} \de_{AB} \sp
g_{\a \m \n} (\r_A)_{\m \n}  = 0 \sp \forall \; A, \a
\end{equation}
is given in App.~\ref{rhosol}. The $\r$'s are proportional to
the $7\times 7$ matrix representation $T^{(\irrep{7})}$ of the $\irrep{7}$ of $\gt$
\begin{subequations}
\begin{equation}
T^{(\irrep{7})}_A \equiv 2 i \r_A \sp
[T_A^{(\irrep{7})},T_B^{(\irrep{7})}] = i f_{ABC} T^{(\irrep{7})}_C
\end{equation}
\begin{equation}
\label{fgt}
f_{ABC} \equiv - 4 {\rm Tr}([\r_A,\r_B] \r_C)
\end{equation}
\end{subequations}
where $f_{ABC}$ are the structure constants of $\gt$.
The second part of \eqref{rhogrel} tells us that the remaining
seven currents in $\so (7)$ can be taken as
\begin{equation}
J_\a (z) \equiv g_{\a \m \n} J_{\m \n} (z) \sp \a = 1 \ldots 7 \ .
\end{equation}
These and the other seven currents $J_{\a 8}(z)$, $\a = 1 \ldots 7$
 of $\so (8)$ transform as $\irrep{7}$'s of $\gt$.

It is convenient to introduce the following linear combinations of
the two $\irrep{7}$'s
\begin{equation}
\label{7combo}
J_\a^\pm (z) \equiv \frac{1}{\sqrt{2}} (J_{\a 8} (z) \pm \frac{i}{2 \sqrt{3}}
J_\a (z) ) \sp \a = 1 \ldots 7
\end{equation}
which also transform as $\irrep{7}$'s under $\gt$.
Then the $\so (8)$ current algebra takes the form
\begin{subequations}
\label{so8d}
\begin{equation}
\label{so8gt}
J_A (z) J_B(w) = \frac{k \de_{AB}}{(z-w)^2}
+ \frac{i f_{ABC} J_C (w)}{z-w} +\Ozw
\end{equation}
\begin{equation}
J_A (z) J_\a^\pm(w) =-\frac{ (T_A^{(\irrep{7})})_{\a \be} J_\be^\pm (w)}{z-w} + \Ozw
= \frac{i f_{A\a \be} J_\be^\pm (w)}{z-w} + \Ozw
\end{equation}
\begin{equation}
\label{pmpm}
J_\a^\pm (z) J_\be^\pm(w) = \pm \sqrt{\frac{2}{3}}
 \frac{g_{\a \be \ga} J_\ga^\mp (w)}{z-w} + \Ozw
\end{equation}
\begin{equation}
\label{pmmp}
J_\a^\pm (z) J_\be^\mp(w) = \frac{k \de_{\a \be}}{(z-w)^2}
+\frac{i f_{\a \be A} J_A (w)}{z-w} + \Ozw
\end{equation}
\end{subequations}
where $g_{\a \be \ga}$ are the octonionic structure constants
in \eqref{oct}, $f_{ABC}$ are the $\gt$ structure constants in
\eqref{fgt} and
\begin{equation}
\label{fAa}
f_{A \a \be} = f_{\a \be A} = - 2(\r_A)_{\a \be} =
 i (T_A^{(\irrep{7})})_{\a \be} \ .
\end{equation}
App.~\ref{rhosol} collects some steps and identities used  in
deriving Eq.~\eqref{so8d}.

The result \eqref{so8d} is the desired diagonal basis for the action
of the triality automorphism $\Tm$:
\begin{subequations}
\label{T1aut}
\begin{equation}
\o_{A,B} = \de_{AB} \sp \o_{\a \pm, \be \pm} = \de_{\a,\be}
e^{\mp \srac{2\pi i}{3}}
\end{equation}
\begin{equation}
 J_A (z)' = J_A (z) \sp J_\a^\pm (z)' = e^{\mp \srac{2\pi i}{3}}
J_\a^\pm (z) \ .
\end{equation}
\begin{equation}
A = 1 \ldots 14 \sp \a = 1 \ldots 7 \ .
\end{equation}
\end{subequations}
In agreement with entry viii) of the table in App.~\ref{appinvsub}, we see
that the triality automorphism $\Tm$ defines the coset space
\begin{equation}
\frac{g}{h} = \frac{\so(8)_x}{(\gt)_x}
\end{equation}
which is not a symmetric space.

\subsection{The triality automorphism $\Tmt$ \label{threeptfive}}

We turn finally to the second triality automorphism $\Tmt$
(see entry ix) of the table in App.~\ref{appinvsub}, which is
inner-automorphically equivalent to the triality automorphism
$\Tm$ but defines the inequivalent coset space
\begin{equation}
\label{T2ss}
\frac{g}{h} = \frac{\so(8)_x}{\su(3)_{3x}}
\end{equation}
with invariant subalgebra $h = \su(3)$ instead of $h=\gt$ for
$\Tm$.
In this discussion, we shall use the standard (Gell-Mann)
Cartesian basis
\begin{equation}
\eta_{AB} = \de_{AB} \sp f_{AB}{}^C = f_{ABC} \sp A,B,C = 1
\ldots 8
\end{equation}
to describe the $ \su (3)$ and, in this basis,
any matrix irrep $T$ of $\su(3)$ satisfies
\begin{subequations}
\label{su3basis}
\begin{equation}
T_A\hc = T_A \sp [T_A,T_B] = if_{ABC} T_C
\end{equation}
\begin{equation}
\bar T_A = -T_A^{\rm t} \sp \bar T_A\hc = \bar T_A \sp
[\bar T_A,\bar T_B] = if_{ABC}\bar T_C
\end{equation}
\end{subequations}
where t is matrix transpose. Note that we have introduced
two equivalent labellings for the same indices $A,B,i,j =
1 \ldots 8$ where $i,j$ are also the vector indices of the
generators $J_{ij}$ of $\so(8)$.

We consider first the $\su(3)$ currents as they are embedded in
the affine $\so(8)$. These may be taken as
\begin{equation}
J_A (z) \equiv - \frac{i}{2} (T_A^{\rm adj})_{ij} J_{ij} (z)
= -\frac{1}{2} f_{Aij} J_{ij}(z) = - \frac{1}{2} f_{ABC} J_{BC}(z)
\end{equation}
where $\{T_A^{\rm adj} \} $ is the adjoint rep of $\su(3)$, and
we find from \eqref{sonca} that these currents satisfy
\begin{subequations}
\begin{equation}
\label{so8su}
J_A (z) J_B(w) = \frac{k_{\rm eff} \de_{AB}}{(z-w)^2}
+ \frac{i f_{ABC} J_C (w)}{z-w} +\Ozw
\end{equation}
\begin{equation}
k_{\rm eff} \equiv k \frac{Q_{\su(3)}}{2} \sp
Q_{\su(3)} \de_{AB} = f_{ACD} f_{BCD} \ .
\end{equation}
\end{subequations}
Then we may compute the invariant level $x_{\su (3)}$ of
the $\su(3)$ currents
\begin{equation}
x_{\su(3)} = \frac{2 k_{\rm eff}}{\psi_{\su(3)}^2} =
k \tilde h_{\su(3)} = 3k  = 3x \quad \Rightarrow
\quad \psi_{\su(3)}^2 = \frac{2}{3} \sp
Q_{\su(3)} = 2 \sp k_{\rm eff} = k
\end{equation}
in agreement with the embedding shown in Eq.~\eqref{T2ss}.

The rest of the $\so(8)$ currents transform as the
$\irrep{ 10}$ and $\overline{\irrep{10}}$ of $h=\su(3)$. We find the
explicit form of these currents as
\begin{subequations}
\label{so8su3}
\begin{equation}
J_{IJK}^\pm (z) = \frac{1}{4} e^{\pm \srac{\pi i}{6}}
(g_{IJK}^\pm)_{ij} J_{ij}(z)
\end{equation}
\begin{equation}
(g_{IJK}^+)_{AB} \equiv (T_A^{(\irrep{3})})_{IL} (T_B^{(\irrep{3})})_{JM}
{\Big \epsilon}_{LMK}
+ \mbox{5 terms} \sp {\Big \epsilon}_{123}=1
\end{equation}
\begin{equation}
(g_{IJK}^-)_{AB}  \equiv (g_{IJK}^+)_{AB}^* =
(\bar T_A^{(\irrep{3})})_{IL} (\bar T_B^{(\irrep{3})})_{JM}
{\Big \epsilon}_{LMK}
+ \mbox{5 terms}
\end{equation}
\begin{equation}
{\rm Tr}(T_A^{\rm adj} g_{IJK}^\pm) =
{\rm Tr}(g_{IJK}^\pm  g_{LMN}^\pm) = 0
\end{equation}
\begin{equation}
A,B,i,j = 1 \ldots 8 \quad ; \qquad  I,J,K,L,M,N = 1 \ldots 3
\quad ; \qquad  \{ IJK \} = 1 \ldots 10
 \end{equation}
\end{subequations}
where ${\Big \epsilon}_{LMK}$ is the Levi-Civita density
and the indices $IJK$ of the tensors $g_{IJK}^\pm$ are
completely symmetrized as indicated.
The matrix irreps $T^{(\irrep{3})}$ and $\bar T^{(\irrep{3})}$ are
the $\irrep{3}$ and $\bar{\irrep{3}}$ of $\su(3)$.

In what follows, we will generally use the composite notation
\begin{equation}
\a = \{ IJK \} = 1 \ldots 10
\end{equation}
for the symmetrized indices. Then we find the following
simple form of the desired diagonal basis for $\Tmt$:
\begin{subequations}
\label{Tmtbas}
\begin{equation}
J_A (z) J_B(w) = \frac{k \de_{AB}}{(z-w)^2}
+ \frac{i f_{ABC} J_C (w)}{z-w} +\Ozw
\end{equation}
\begin{equation}
J_A (z) J_\a^+(w) =-
 \frac{(T_A^{(\irrep{10})})_{\a \be}   J_\be^+(w)}{z-w} + \Ozw
\end{equation}
\begin{equation}
J_A (z) J_\a^-(w) =-
 \frac{(\bar T_A^{(\irrep{10})})_{\a \be}   J_\be^-(w)}{z-w} + \Ozw
\end{equation}
\begin{equation}
J_\a^+ (z) J_\be^-(w) =\frac{k (\tone)_{\a \be}}{(z-w)^2}
-  \frac{(T_A^{(\irrep{10})})_{\a \be} J_A (w)}{z-w} + \Ozw
\end{equation}
\begin{equation}
J_\a^- (z) J_\be^+(w) =\frac{k (\tone)_{\a \be}}{(z-w)^2}
-  \frac{(\bar T_A^{(\irrep{10})})_{\a \be} J_A (w)}{z-w} + \Ozw
\end{equation}
\begin{equation}
J_\a^\pm (z) J_\be^\mp(w) = \pm
\frac{\tilde g_{\a \be \ga} J_\ga^\pm (w)}{z-w} + \Ozw \ .
\end{equation}
\end{subequations}
Here we have defined the $10\times 10$ matrices
\begin{subequations}
\begin{equation}
(T_A^{(\irrep{10})})_{IJK,LMN} \equiv
\frac{1}{12} \left\{ (T_A^{(\irrep{3})})_{IL} \de_{JM} \de_{KN}
+ \mbox{35 terms} \right\}
\end{equation}
\begin{equation}
(\bar T_A^{(\irrep{10})})_{IJK,LMN} \equiv
-(T_A^{(\irrep{10})})_{LMN,IJK} =
\frac{1}{12} \left\{ (\bar T_A^{(\irrep{3})})_{IL} \de_{JM} \de_{KN}
+ \mbox{35 terms} \right\}
\end{equation}
\end{subequations}
which are the $\irrep{10}$ and $\overline{\irrep{10}}$ irreps
of $\su(3)$, and the additional tensors
\begin{subequations}
\begin{equation}
\label{gtensor}
\tilde g_{IJK,LMN,PQR} \equiv \frac{1}{216} (
{\Big \epsilon}_{ILP} {\Big \epsilon}_{JMQ} {\Big \epsilon}_{KNR} +
\mbox{215 terms} )
\end{equation}
\begin{equation}
\label{tone}
(\tone)_{IJK,LMN} \equiv \frac{1}{36} (\de_{IL} \de_{JM} \de_{KN}
+\mbox{35 terms} ) \ .
\end{equation}
\end{subequations}
Useful identities among all these matrices and tensors are
collected in App.~\ref{appT2}, where it is also noted that the matrix
$\tone$ in \eqref{tone} is the natural unit matrix in the $10 \times 10$
space.

In the basis \eqref{Tmtbas}, the action of the triality
automorphism $\Tmt$ has the same diagonal form
\begin{subequations}
\begin{equation}
J_A(z)' = J_A (z)\sp J_\a^{\pm}(z)' = e^{\mp \srac{2\pi i}{3}}
J_\a^{\pm} (z)
\end{equation}
\begin{equation}
A = 1 \ldots 8 \sp \a = 1 \ldots 10
\end{equation}
\end{subequations}
 as that given for $\Tm$ in \eqref{T1aut}.

\subsection{The affine-Sugawara constructions}

In the Cartesian basis \eqref{cartbas},
the affine-Sugawara construction
\cite{Bardakci:1971nb,Halpern:1971ay,Dashen:1975hp,Knizhnik:1984nr,Segal}
for level $k=x$ of $\so (2n)$ is well-known
\begin{equation}
T(z) =
\frac{1}{2k + Q} : \sum_{i < j} J_{ij} (z) J_{ij} (z) : \sp
Q = 4 (n-1) \sp c = \frac{x n (2n -1)}{x + 2(n-1)}
\end{equation}
and this construction is easily rewritten in the diagonal bases
above: In any basis, the general form of the affine-Sugawara
construction  on simple $g$ is
\begin{subequations}
\begin{equation}
\label{Tut}
T(z) = L_g^{ab} : J_a(z) J_b(z) : \sp L_g^{ab} =
\frac{\eta^{ab}}{2k + Q_g}
\end{equation}
\begin{equation}
J_a (z) J_b (w) = \frac{k\eta_{ab}}{(z-w)^2} + \frac{if_{ab}{}^c J_c(w)}
{z-w} + \Ozw \sp a,b = 1 \ldots {\rm dim}\,g
\end{equation}
\end{subequations}
so that the Killing metric $\eta_{ab}$ and its inverse $\eta^{ab}$
are easily read from the current-current OPEs in any basis.

We find in particular for the diagonal bases of each of our examples: \nl
\ni {\bf $\Pm$}
\begin{equation}
\label{Tpar}
T(z) = \frac{1}{2k + Q} : \sum_{\m < \n} J_{\m \n} (z) J_{\m \n} (z)
+ \sum_{\m} J_\m (z) J_\m (z) :
\end{equation}
\vskip .3cm
\ni {\bf $\Cmg$}
\begin{equation}
\label{Tcha}
T(z) = \frac{1}{2k + Q} : \sum_{A < B} J_{AB} (z) J_{AB} (z) +
\sum_{I < J} J_{IJ} (z) J_{IJ} (z)
+ \sum_{A,I} J_{AI} (z) J_{AI} (z) :
\end{equation}
\vskip .3cm
\ni {\bf $\Tm$} and {\bf $\Tmt$}
\begin{equation}
\label{Ttr}
T(z) =
  \frac{1}{2k + Q} : \sum_{A} J_{A} (z) J_{A} (z)
+ \sum_{\a} \Big(J_{\a}^+ (z) J_{\a}^- (z)+ J_{\a}^- (z) J_{\a}^+ (z)
\Big): \ .
\end{equation}
The form \eqref{Ttr} for the triality automorphisms can also be
checked directly using the sum rules in \eqref{sumr1} and
\eqref{sumr2}.
It is easy to check that each affine-Sugawara construction in this
list is invariant
\begin{equation}
T(z)' = L_g^{ab} : J_a (z)' J_b (z)' : \ = T(z)
\end{equation}
under the diagonal action of each of the automorphisms discussed above.

\section{The Twisted Sectors $\Pmo$, $\Cmgo$ and
$\Tmo$, $\Tmto$}

\subsection{The twisted current algebras}

In current-algebraic orbifold theory
\cite{deBoer:1999na,Evslin:1999ve,Halpern:2000vj,deBoer:2001nw,Halpern:2002ww,Halpern:2002zv,Halpern:2002vh},
 the twisted OPEs and
monodromies of the twisted sectors of each orbifold $A(H)/H$
are obtained from the OPEs and automorphic responses of the
untwisted sector $A(H)$ by the method of eigenfields and the principle of local isomorphisms
\cite{Borisov:1997nc,deBoer:1999na,Halpern:2000vj,deBoer:2001nw}. As a simple
example, the monodromy of the general twisted left-mover currents
$\hj$ of sector $\s$ is
\begin{subequations}
\begin{equation}
\label{genmon}
\hj_\nrm (z e^{2\pi i},\s) = E_{n(r)}(\s)\hj_\nrm
(z,\s) \sp
E_{n(r)}(\s) =
e^{- 2 \pi i \srac{n(r)}{\r(\s)}}
\end{equation}
\begin{equation}
\hj_{n(r) \pm \r(\s),\m} (z,\s) = \hj_\nrm (z,\s)
\end{equation}
\begin{equation}
\label{nbarr}
 \bar n(r) \in \{ 0 , \ldots , \r(\s) -1 \} \sp
 \srange
 \end{equation}
\end{subequations}
where $\r (\s)$ is the order $h_\s \in H$, $N_c$ is the number
of conjugacy classes of $H$  and the spectral
indices $\{ n(r)\}$ are determined from the $H$-eigenvalue
problem \cite{deBoer:1999na,Halpern:2000vj,deBoer:2001nw}
\begin{equation}
\w (h_\s) U\hc(\s)^\nrm = U\hc(\s)^\nrm E_{n(r)} (\s) \ .
\end{equation}
Here $\o (h_\s)$ is the action of $h_\s\in H$ on the untwisted currents,
$\{ \m \}$ are the degeneracy indices of the eigenvalue problem and
the integers $\bar n(r)$ in \eqref{nbarr} are the pullback of the spectral
indices to the fundamental range.

Solving the $H$-eigenvalue problem is equivalent to finding a diagonal basis
for the automorphism, and since we have already found such diagonal
bases, we may now choose the trivial solution
\begin{equation}
\label{trivsol}
 U\hc(\s) = \one \sp E_{n(r)}
= {\rm diag}\,\w (h_\s)
\end{equation}
from which the spectral indices are easily deduced.
Then we obtain in particular \nl
\ni {\bf $\Pmo$}
\begin{subequations}
\begin{equation}
\r = 2 \sp \bar n_{\m \n} =0 \sp \bar n_\m =1
\end{equation}
\begin{equation}
\hj_{0,\m \n} (z e^{2 \pi i}) = \hj_{0,\m \n} (z) \sp
\hj_{1,\m } (z e^{2 \pi i}) = -\hj_{1,\m } (z)
\end{equation}
\end{subequations}
\vskip .3cm
\ni {\bf $\Cmgo$}
\begin{subequations}
\begin{equation}
\r = 2 \sp \bar n_{AB} =\bar n_{IJ} = 0 \sp \bar n_{AI} =1
\end{equation}
\begin{equation}
\hj_{0,AB} (z e^{2\pi i}) = \hj_{0,AB} (z) \sp
\hj_{0,IJ} (z e^{2\pi i}) = \hj_{0,IJ} (z) \sp
\hj_{1,AI} (z e^{2\pi i}) = -\hj_{1,AI} (z)
\end{equation}
\end{subequations}
\vskip .3cm
\ni {\bf $\Tmo$} and  {\bf $\Tmto$}
\begin{subequations}
\begin{equation}
\r = 3 \sp \bar n_{A}  = 0 \sp \bar n_{+1,\a} =1 \sp
\bar n_{-1,\a} =2
\end{equation}
\begin{equation}
\hj_{0,A} (z e^{2\pi i}) = \hj_{0,A} (z) \sp
\hj_{\pm 1,\a} (z e^{2\pi i}) = e^{\mp \srac{2 \pi i}{3}}
\hj_{\pm 1,\a} (z) 
\end{equation}
\end{subequations}
for the twisted sectors $\so (2n)/H$ of the orbifolds on $\so(2n)$.

Returning to the general case, the monodromy \eqref{genmon}
gives the expansion
\begin{subequations}
\begin{equation}
\hj_\nrm (z,\s) = \sum_{m \in \Zm} \hj_\nrm (m + \srac{n(r)}{\r(\s)})
z^{-(m + \srac{n(r)}{\r(\s)})-1}
\end{equation}
\begin{equation}
\label{period}
\hj_{n(r) \pm \r(\s)} (m  + \srac{n(r) \pm \r(\s)}{\r(\s)})
= \hj_\nrm (m \pm 1 + \srac{n(r)}{\r(\s)})
\end{equation}
\end{subequations}
where $\hj_\nrm (m + \srac{n(r)}{\r(\s)})$ are the general
twisted current modes. This gives in particular for our examples: \nl
{\bf $\Pmo$}
\begin{equation}
\hj_{0,\m \n} (z ) = \sum_m \hj_{0,\m\n}(m) z^{-m-1} \sp
\hj_{1,\m } (z) = \sum_m \hj_{1,\m}(m +\srac{1}{2}) z^{-(m +\srac{1}{2})-1}
\end{equation}
\skl
{\bf $\Cmgo$}
\begin{subequations}
\begin{equation}
\hj_{0,AB} (z) = \sum_{m} \hj_{0,AB} (m) z^{-m-1} \sp
\hj_{0,IJ} (z) = \sum_{m} \hj_{0,IJ} (m) z^{-m-1}
\end{equation}
\begin{equation}
\hj_{1,AI} (z) = \sum_{m} \hj_{1,AI} (m +\srac{1}{2}) z^{-(m +\srac{1}{2})-1}
\end{equation}
\end{subequations}
\skl {\bf $\Tmo$} and {\bf $\Tmto$}
\begin{equation}
\hj_{0,A} (z) = \sum_{m} \hj_{0,A} (m) z^{-m-1} \sp
\hj_{\pm 1,\a} (z) = \sum_{m} \hj_{\pm 1,\a} (m \pm \srac{1}{3})
z^{-(m \pm \srac{1}{3})-1} \ .
\end{equation}
We also give the examples of mode periodicity
\begin{subequations}
\begin{equation}
\hj_{2,AB} (m + \srac{2}{2} ) \!=\! \hj_{0,AB} (m +1) ,\,\,\,\,
%\end{equation}
%\begin{equation}
\hj_{2,IJ} (m + \srac{2}{2} ) \!=\! \hj_{0,IJ} (m +1) ,\,\,\,\,
%\end{equation}
%\begin{equation}
\hj_{-1,AI} (m - \srac{1}{2} ) \!=\! \hj_{1,AI} (m -1 + \srac{1}{2})
\end{equation}
\begin{equation}
\hj_{\pm 2,\a} ( m \pm \srac{2}{3}) = \hj_{\mp 1,\a} (m \pm1 \mp \srac{1}{3})
\end{equation}
\end{subequations}
which follow from the general form in Eq.~\eqref{period}.

For any current-algebraic orbifold $A(H)/H$, the general twisted current algebra of sector $\s$ is
\cite{Halpern:2000vj,deBoer:2001nw}
\begin{subequations}
\label{malg0}
\begin{eqnarray}
\label{malg}
 [\hj_\nrm(\mmrrs),\hj_\nsn(\nnsrs)] & = &
i \F_{\nrm, \nsn}{}^{n(r) + n(s),\de} (\s) \hj_{n(r)+n(s),\de}
( m + n + \srac{n(r) + n(s)}{\r (\s)}) \nn \\
 & &+   (\mmrrs)\de_{\mnnrnsrsf,0}\sG_{\nrm;\mnrn}(\s) 
\end{eqnarray}
\begin{equation}
\s = 0 , \ldots , N_c -1
\end{equation}
\end{subequations}
where the zero modes $\{ \hj_{0\m}(0)\}$ generate the residual
symmetry algebra and Ward identities of sector $\s$.
General formulas are given for the twisted structure constants
 $\F (\s)$ and the twisted metric $\sG(\s)$ in
 Refs.~\cite{deBoer:1999na,Halpern:2000vj}, but these reduce
 to $\G = k \eta$ and $\F =f$ for trivial normalization
$\chi =1$ and the trivial solution \eqref{trivsol} of the
$H$-eigenvalue problem.

This gives in particular the outer-automorphically twisted current
algebra for the twisted sector $\Pmo$:
\begin{subequations}
\begin{eqnarray}
[ \hj_{0,\m \n}(m),\hj_{0,\r \s} (n)]
& = &   i (\de_{\n \r} \hj_{0,\m \s}(m+n) + \de_{\m \s} \hj_{0,\n \r} (m+n) \nn \\
&&  -\de_{\m \r} \hj_{0,\n \s} (m+n)- \de_{\n \s} \hj_{0,\m \r}(m+n) ) \nn \\
&& + 2k (e_{\m \n})_{\r \s} m \de_{m+n,0} \\
{}[ {\hj}_{0,\m \n}(m ),\hj_{1, \r} (n+\srac{1}{2})]
 &= &  i (\de_{\n \r} \hj_{1, \m}(m+n +\srac{1}{2}) - \de_{\m \s}
 \hj_{1,\n} (m+n+\srac{1}{2}) ) \\
{}[ {\hj}_{1,\m}(m +\srac{1}{2}),\hj_{1,\n} (n+\srac{1}{2})]
%& = &  - i \hj_{2,\m \n}(m+n +\srac{2}{2}) + k \de_{\m \n}
% (m + \srac{1}{2}) \de_{m+n+1,0} \\
 &=&   - i \hj_{0,\m \n}(m+n +1) + k \de_{\m \n}
 (m + \srac{1}{2}) \de_{m+n+1,0}
 \end{eqnarray}
\begin{equation}
\hj_{0,\m\n} (m) \hc = \hj_{0,\m\n} (-m) \sp
\hj_{1,\m} (m + \srac{1}{2}) \hc = \hj_{-1,\m} (-m-\srac{1}{2})
= \hj_{1,\m} (-m-1 + \srac{1}{2})
\end{equation}
\begin{equation}
\{ \m, \n , \r, \s \} = \{ 1, \ldots, 2n -1\}
\end{equation}
\end{subequations}
and for the twisted sectors $\Cmgo$:
\begin{subequations}
\begin{eqnarray}
[ \hj_{0,AB}(m),\hj_{0,CD} (n)]
& = &   i (\de_{BC} \hj_{0,AD}(m+n) + \de_{AD} \hj_{0,BC} (m+n) \nn \\
&&  -\de_{AC} \hj_{0,BD} (m+n)- \de_{BD} \hj_{0,AC}(m+n) ) \nn \\
&& + 2k (e_{AB})_{CD} m \de_{m+n,0}
\end{eqnarray}
\begin{eqnarray}
[ \hj_{0,IJ}(m),\hj_{0,KL} (n)]
& = &   i (\de_{JK} \hj_{0,IL}(m+n) + \de_{IL} \hj_{0,JK} (m+n) \nn \\
&&  -\de_{IK} \hj_{0,JL} (m+n)- \de_{JL} \hj_{0,IK}(m+n) ) \nn \\
&& + 2k (e_{IJ})_{KL} m \de_{m+n,0}
\end{eqnarray}
\begin{equation}
[ \hj_{0,AB}(m),\hj_{0,IJ} (n)] = 0
\end{equation}
\begin{eqnarray}
[ \hj_{1,AI}(m +\srac{1}{2}),\hj_{1,BJ} (n+\srac{1}{2})]
& = &  - i (\de_{AB} \hj_{0,IJ}(m+n +1) + \de_{IJ} \hj_{0,AB} (m+n+1) \nn) \\
&& + k \de_{AB} \de_{CD} (m  +\srac{1}{2}) \de_{m+n+1,0}
\end{eqnarray}
\begin{equation}
[ \hj_{0,AB}(m ),\hj_{1,CI} (n+\srac{1}{2})]
 =   i (\de_{BC} \hj_{1,AI}(m+n +\srac{1}{2}) - \de_{AC} \hj_{1,BI} (m+n+\srac{1}{2}) \nn)
\end{equation}
\begin{equation}
[ \hj_{0,IJ}(m ),\hj_{1,AK} (n+\srac{1}{2})]
 =   i (\de_{JK} \hj_{1,AI}(m+n +\srac{1}{2}) - \de_{IK} \hj_{1,AJ} (m+n+\srac{1}{2}))
\end{equation}
\begin{equation}
\hj_{0,AB} (m) \hc = \hj_{0,AB} (-m) \sp
\hj_{0,IJ} (m) \hc = \hj_{0,IJ} (-m)
\end{equation}
\begin{equation}
\hj_{1,AI} (m + \srac{1}{2}) \hc = \hj_{-1,AI} (-m-\srac{1}{2})
= \hj_{1,AI} (-m-1 + \srac{1}{2})
\end{equation}
\begin{equation}
\{ A, B , C, D \} = \{ 1 \ldots r\} \sp
\{ I, J , K, L \} = \{ r+1, \ldots , 2n\} \sp r  = n, \ldots ,
2n-1
\end{equation}
\end{subequations}
and for the twisted triality sector $\Tmo$:
\begin{subequations}
\label{tralg}
\begin{equation}
[ \hj_{0,A}(m),\hj_{0,B} (n)] = if_{ABC} \hj_{0,C}(m+n)
+ k \de_{AB} m \de_{m+n,0}
\end{equation}
\begin{equation}
[ \hj_{0,A}(m),\hj_{\pm 1,\a} (n \pm \srac{1}{3} )] =
if_{A \a \be} \hj_{\pm 1,\be}(m+n \pm \srac{1}{3} )
\end{equation}
\begin{equation}
\label{tralgc}
[ \hj_{\pm 1,\a}(m \pm \srac{1}{3}),
\hj_{\pm 1,\be} (n \pm \srac{1}{3} )] =
\pm \sqrt{\frac{2}{3}}
g_{\a \be \ga} \hj_{\pm 2,\ga}(m+n \pm  \srac{2}{3} )=
\pm \sqrt{\frac{2}{3}}
g_{\a \be \ga} \hj_{\mp 1,\ga}(m+n \pm 1 \mp \srac{1}{3} )
\end{equation}
\begin{equation}
[ \hj_{\pm 1,\a}(m \pm \srac{1}{3}),
\hj_{\mp 1,\be} (n \mp \srac{1}{3} )] =
i f_{\a \be A} \hj_{0,A} (m +n)  + k \de_{\a \be}(m\pm \srac{1}{3} )
\de_{m+n,0}
\end{equation}
\begin{equation}
\hj_{0,A} (m) \hc = \hj_{0,A} (m) \sp
\hj_{\pm 1,\a} (m \pm \srac{1}{3}) \hc =
\hj_{\mp 1,\a} (-m \mp \srac{1}{3})
\end{equation}
\begin{equation}
\{ A, B , C \} = \{ 1  \ldots 14 \} \sp
\{ \a, \be , \ga \} = \{ 1  \ldots 7 \}
\end{equation}
\end{subequations}
and finally for the twisted triality sector $ \Tmto$:
\begin{subequations}
\label{tralg2}
\begin{equation}
[ \hj_{0,A}(m),\hj_{0,B} (n)] = if_{ABC} \hj_{0,C}(m+n)
+ k \de_{AB} m \de_{m+n,0}
\end{equation}
\begin{equation}
[ \hj_{0,A}(m),\hj_{+1,\a} (n + \srac{1}{3} )] =-
(T_A^{(\irrep{10})})_{\a \be}
 \hj_{+ 1,\be}(m+n + \srac{1}{3} )
\end{equation}
\begin{equation}
[ \hj_{0,A}(m),\hj_{-1,\a} (n - \srac{1}{3} )] =-
(\bar T_A^{(\irrep{10})})_{\a \be}
 \hj_{- 1,\be}(m+n - \srac{1}{3} )
\end{equation}
\begin{equation}
[ \hj_{+1,\a}(m + \srac{1}{3}),
\hj_{-1,\be} (n - \srac{1}{3} )] =
-(T_A^{(\irrep{10})})_{\a \be} \hj_{0,A} (m+n) +
k (\tone)_{\a \be}(m + \srac{1}{3} ) \de_{m+n,0}
\end{equation}
\begin{equation}
[ \hj_{-1,\a}(m - \srac{1}{3}),
\hj_{+1,\be} (n + \srac{1}{3} )] =
-(\bar T_A^{(\irrep{10})})_{\a \be} \hj_{0,A} (m+n) +
k (\tone)_{\a \be}(m - \srac{1}{3} ) \de_{m+n,0}
\end{equation}
\begin{equation}
[ \hj_{\pm 1,\a}(m \pm \srac{1}{3}),
\hj_{\pm 1,\be} (n \pm \srac{1}{3} )] = \pm
\tilde g_{\a \be \ga} \hj_{\mp 1,\ga}(m +n \pm 1 \mp\srac{1}{3} )
\end{equation}
\begin{equation}
\hj_{0,A} (m) \hc = \hj_{0,A} (m) \sp
\hj_{\pm 1,\a} (m \pm \srac{1}{3}) \hc =
\hj_{\mp 1,\a} (-m \mp \srac{1}{3})
\end{equation}
\begin{equation}
\{ A, B , C \} = \{ 1  \ldots 8 \} \sp
\{ \a, \be , \ga \} = \{ 1  \ldots 10 \} \ .
\end{equation}
\end{subequations}
The twisted current algebras of the sectors
$\so (8)/\Tm^2$ and $\so(8)/\Tmt^2$ are discussed in
Subsec.~\ref{sectri}.

\subsection{Rectification \label{recsec} }

In current-algebraic orbifold theory, it is known
\cite{deBoer:2001nw} that the twisted right-mover current
algebra
\begin{eqnarray}
\label{malgr}
 [\hjb_\nrm(\mmrrs),\hjb_\nsn(\nnsrs)] & = &
i \F_{\nrm, \nsn}{}^{n(r) + n(s),\de} (\s) \hjb_{n(r)+n(s),\de}
( m + n + \srac{n(r) + n(s)}{\r (\s)}) \nn \\
 & &-   (\mmrrs)\de_{\mnnrnsrsf,0}\sG_{\nrm;\mnrn}(\s)
\end{eqnarray}
of sector $\s \leftrightarrow h_\s \in H$
is the same as the twisted left-mover current algebra \eqref{malg0},
but with the {\it sign reversal} of the central term shown here.
As discussed in Ref.~\cite{deBoer:2001nw}, the twisted right-mover 
current algebra \eqref{malgr} is isomorphic to the twisted left-mover 
current algebra of sector $h_\s^{-1}$ and, as a consequence, the form 
\eqref{malgr} is in agreement\footnote{The bracket analogue of the 
twisted right-mover current algebra \eqref{malgr} also follows from 
the general WZW orbifold action \cite{deBoer:2001nw}.}
with earlier analysis at the level of characters \cite{Dijkgraaf:1989hb}.

Nevertheless, it has been found on a case-by-case basis that each
of the twisted right-mover current algebras so far considered
can be {\it rectified} \cite{deBoer:2001nw} by a linear
transformation into
a copy  $\hjbb$ of the twisted left-mover current algebra: \nl
$\bullet$ the WZW permutation orbifolds
\cite{deBoer:2001nw,Halpern:2002ww,Halpern:2002vh} \nl
$\bullet$ the inner-automorphic WZW orbifolds
\cite{deBoer:2001nw} on simple $g$ \nl
$\bullet$ the (outer-automorphic) charge conjugation
orbifold on $\su (n\geq 3)$ \cite{Halpern:2002ww}. \nl
So far as the basic types of twisted right-mover current algebras
are concerned,
this leaves the rectification problem open only for the other
outer-automorphically twisted affine Lie algebras on simple $g$.

As a simple example of rectification, we consider first the general inner- or
outer-automorphic WZW orbifold of $\Zm_2$-type,
whose twisted left-mover current algebra has the general form
\begin{subequations}
\label{genz2alg}
\begin{equation}
[ \hj_{0,A}(m),\hj_{0,B} (n)] = i\F_{0,A;0,B}{}^{0,C} \hj_{0,C}(m+n)
+  \sG_{0,A;0,B} m \de_{m+n,0}
\end{equation}
\begin{equation}
[ \hj_{0,A}(m),\hj_{1,I} (n + \srac{1}{2} )] =
i\F_{0,A ;1,I}{}^{1,J} \hj_{1,J}(m+n + \srac{1}{2} )
\end{equation}
\begin{equation}
[ \hj_{1,I}(m + \srac{1}{2}),
\hj_{1,J} (n + \srac{1}{2} )] = i \F_{1,I;1,J}{}^{0,A}
\hj_{0,A}(m+n +1) +  \sG_{1,I;1,J} (m + \srac{1}{2})\de_{m+n+1,0}
\end{equation}
\begin{equation}
A,B,C \in h \sp I,J \in g/h
\end{equation}
\end{subequations}
with $g/h$ a symmetric space. This form includes in particular
the $\Zm_2$-twisted current algebras of all the special cases
\begin{equation}
\Cmgo \supset \Cmo \ \ldots \ \Pmo
\end{equation}
given above. Then, reversing the signs
of the central terms to obtain the corresponding twisted
right-mover current algebra, we find that the following redefinition
\begin{equation}
\label{z2rect}
\hjbb_{0,A} (m) \equiv \hat{\bar{J}}_{0,A}(-m) \sp
\hjbb_{1,I} (m + \srac{1}{2}) \equiv
\hat{\bar{J}}_{-1,I}(-m-\srac{1}{2})
= \hjb_{1,I}(-m-1 + \srac{1}{2})
\end{equation}
rectifies the twisted right-mover current algebra into a copy of the
twisted left-mover current algebra \eqref{genz2alg}. In fact, there is a theorem
\cite{deBoer:2001nw} which tells us that the twisted right-mover
currents of all $\Zm_2$-type orbifolds are
rectifiable because $h_\s^{-1} = h_\s$ for the non-trivial
element of the $\Zm_2$. The result \eqref{z2rect} shows
explicitly that this can be done without extra phases.

On the other hand, we find from Eq.~\eqref{tralg} that some
non-trivial phases are necessary
%{\bf $\Tmo$}
\begin{equation}
\label{recT1}
\hjbb_{0,A} (m ) = \hjb_{0,A} (-m) \sp
\hjbb_{\pm 1,\a}(m \pm \srac{1}{3}) =
 - \hjb_{\mp 1,\a} ( - m \mp \srac{1}{3})
\end{equation}
to rectify the twisted right-mover current algebra of the
twisted triality sector $\so(8)/\Tm$.

We have also been able to find the rectification for the twisted
triality sector $\so(8)/\Tmt$:
\begin{equation}
\label{recT2}
\hjbb_{0,A} (m ) = \omega_{AB} \hjb_{0,B} (-m) \sp
\hjbb_{\pm 1,\a}(m \pm \srac{1}{3}) =
 - \hjb_{\mp 1,\a} ( - m \mp \srac{1}{3}) \ .
\end{equation}
In this case $\omega_{AB}$ cannot be trivial and the simplest
choice is $\omega = \omega (\Cm)$, the (outer-automorphic) action
of charge conjugation $\Cm$ on $\su(3)$:
\begin{subequations}
\begin{equation}
T_A^{(\irrep{3})}{}' = \omega_{AB} T_B^{(\irrep{3})} =
\bar T_A^{(\irrep{3})} \sp
\bar T_A^{(\irrep{3})}{}'=\omega_{AB} \bar T_B^{(\irrep{3})} =
T_A^{(\irrep{3})}
\end{equation}
\begin{equation}
\Rightarrow \quad
T_A^{(\irrep{10})}{}' = \omega_{AB} T_B^{(\irrep{10})} =
\bar T_A^{(\irrep{10})} \sp
\bar T_A^{(\irrep{10})}{}'=\omega_{AB} \bar T_B^{(\irrep{10})} =
T_A^{(\irrep{10})}  \ .
\end{equation}
\end{subequations}
The diagonal action of this automorphism in this Cartesian basis
is given in Ref.~\cite{Halpern:2002ww}, where it is noted that
the symmetric space
\begin{equation}
\frac{g}{h} = \frac{\su(3)_{3x}}{\so(3)_{12x}}
\end{equation}
is defined by this automorphism. As discussed in Subsec.~\ref{sectri},
the rectifications \eqref{recT1} and \eqref{recT2} hold as well for
the twisted triality sectors $\so(8)/\Tm^2$ and $\so(8)/\Tmt^2$
respectively.

Together with the conclusions of
Refs.~\cite{deBoer:2001nw,Halpern:2002ww,Halpern:2002vh},
this completes the rectification of all the basic types
of twisted right-mover current algebras. The rectification
problem is not completely solved however because there exist
more general twisted current algebras associated to the
composition of automorphisms of  different basic types.
Examples of these are the ``doubly-twisted'' current
algebras of Refs.~\cite{Evslin:1999qb,Evslin:1999ve}, which result
from the composition of permutations of copies of $\mathfrak{g}$
with inner automorphisms of $\mathfrak{g}$.

\subsection{The twisted affine-Sugawara constructions \label{twassec}}

For any current-algebraic orbifold $A(H)/H$, the
{\it twisted affine-Virasoro construction} of sector $\s$ is
\cite{deBoer:1999na,Halpern:2000vj}
\begin{equation}
\label{twvir}
\hat T_\s (z) ={\cal{L}}^{\nrm ; -\nrn} (\s) : \hj_\nrm (z,\s)
\hj_{-\nrn} (z,\s):
\sp \hat c =  c
 \end{equation}
 where ${\cal{L}} (\s)$ is the twisted inverse inertia tensor
 and $ : \cdot :$ is operator product normal ordering
\cite{Evslin:1999qb,deBoer:1999na,Halpern:2000vj} of the twisted currents. The explicit form
\cite{deBoer:1999na,Halpern:2000vj} of ${\cal{L}} (\s) = {\cal{L}} (L,\s)$
for all $A(H)/H$ is a duality transformation of the inverse
inertia tensor $L^{ab}$ of the corresponding untwisted
affine-Virasoro construction
\cite{Halpern:1989ss,Morozov:1990uu,Halpern:1992gb,Halpern:1996js} of the
symmetric CFT $A(H)$.
The special case of the WZW orbifolds is described by the general
{\it twisted affine-Sugawara construction} \cite{deBoer:1999na,Halpern:2000vj}
\begin{subequations}
\label{twasc}
\begin{equation}
\hat T_\s (z) ={\lr}^{\nrm ; -\nrn} (\s) : \hj_\nrm (z,\s)
\hj_{-\nrn} (z,\s):
\sp \hat c_g=  c_g
 \end{equation}
\begin{equation}
\lr (\s) = {\cal{L}} (L_g,\s)
\end{equation}
\end{subequations}
where $L_g^{ab}$  is the ordinary inverse inertia tensor of the untwisted
affine-Sugawara construction
\cite{Bardakci:1971nb,Halpern:1971ay,Dashen:1975hp,Knizhnik:1984nr,Segal}
$T = L_g^{ab} : J_a J_b:$ in the symmetric theory $A_g(H)$.

Because each of our
automorphisms acts in a diagonal basis, the twisted inverse
inertia tensor of each of our twisted sectors is equal to the
ordinary inverse inertia tensor
$\lr (\s) = L_g$. Then Eqs.~\eqref{Tpar}-\eqref{Ttr} and
\eqref{twasc} give the explicit forms of the
twisted affine-Sugawara constructions \vskip .3cm %\nl
\ni {\bf $\Pmo$}
\begin{equation}
\hat T (z) = \frac{1}{2k + Q}
: \sum_{\m < \n} \hj_{0,\m \n} (z)\hj_{0,\m \n }(z)
 + \sum_{\m} \hj_{1,\m } (z) \hj_{-1,\m }(z) :
\end{equation}
\skl {\bf $\Cmgo$}
\begin{equation}
\hat T (z) = \frac{1}{2 k + Q}
: \sum_{A < B} \hj_{0,AB} (z) \hj_{0,AB} (z)
+
\sum_{I < J} \hj_{0,IJ} (z) \hj_{0,IJ} (z)
+
 \sum_{A,I} \hj_{1,AI} (z) \hj_{-1,AI} (z) :
\end{equation}
\skl {\bf $\Tmo$} and {\bf $\Tmto$}
\begin{equation}
\label{tras}
\hat T (z) = \frac{1}{2 k + Q}
: \sum_{A} \hj_{0,A} (z) \hj_{0,A} (z)
+ \sum_{\a} (\hj_{1,\a} (z) \hj_{-1,\a} (z)+ \hj_{-1,\a} (z)
\hj_{1,\a} (z)):
\end{equation}
for each of our examples.

We turn next to the left- and right-mover conformal weights
$\hat{\Delta}_0 (\s)  $, $\hat{\bar{\Delta}}_0 (\s) $
of the scalar twist-field state of sector $\s$, whose general form is
\begin{subequations}
\label{gcw}
\begin{equation}
\hat T_\s (z) = \sum_m L_\s (m) z^{-m-2} \sp
L_\s (m \geq 0) | 0 \rangle_\s = \de_{m,0} \hat \Delta_0 (\s)
| 0 \rangle_\s
\end{equation}
\begin{equation}
\label{Del0vir}
\hat{\Delta}_0 (\s)  =
 \sum_{r,\m,\n} {\cal{L}}^{\nrm ; \mnrn} (\s)
\frac{\nb}{2 \rho (\s)} \left( 1 - \frac{\nb}{\rho (\s)} \right)
 \G_{\nrm ; \mnrn} (\s) = \hat{\bar{\Delta}}_0 (\s)
\end{equation}
\end{subequations}
in each sector of every current-algebraic orbifold $A(H)/H$.
The result \eqref{Del0vir} is most easily derived
by going over to the $M$ or mode-ordered form
\cite{Halpern:2000vj,deBoer:2001nw}
of the Virasoro generators and using \eqref{vacl}, \eqref{vacr}.
The special case of \eqref{Del0vir} for the WZW orbifolds
\begin{equation}
\label{Del0}
\hat{\Delta}_0 (\s)  =
 \sum_{r,\m,\n} \lr^{\nrm ; \mnrn} (\s)
\frac{\nb}{2 \rho (\s)} \left( 1 - \frac{\nb}{\rho (\s)} \right)
 \G_{\nrm ; \mnrn} (\s)
\end{equation}
is not difficult to evaluate for particular classes of WZW orbifolds
(see Refs.~\cite{deBoer:2001nw,Halpern:2002ww,Halpern:2002vh}
for the WZW permutation orbifolds, Ref.~\cite{Halpern:2002ww}
for the outer-automorphic charge conjugation orbifold on
$\su (n)$ and Ref.~\cite{Halpern:2002vh}
for the inner-automorphic WZW orbifolds).

For our purposes, we note that the general result \eqref{Del0}
is also easy to evaluate for all
low-order inner or outer automorphisms of the large class of
permutation-invariant Lie algebras $g$
\begin{subequations}
\begin{equation}
\label{gsum}
g = \oplus \, \gm^I \sp \gm^I \simeq \gm \sp k_I = k
\end{equation}
\begin{equation}
\label{Delsp}
\hat \Delta_0 (\s) = \epsilon \,{\rm dim}\, g/h\frac{x_\gm}{x_\gm + \tilde h_\gm}
\sp \epsilon = \left\{ \begin{array}{ll}
\frac{1}{16} & \mbox{for}\;\, \r =2 \\
\frac{1}{18} & \mbox{for}\;\, \r =3 \ .
\end{array}\right.
\end{equation}
\end{subequations}
Here $h$ is the invariant subalgebra of $g$, while $\tilde h_\gm$ and
$x_\gm$ are respectively the dual Coxeter number of $\gm$ and the
invariant level of affine $\gm$. To see this for order $\r =2$ and
simple $g$, use the identities
\begin{subequations}
\begin{equation}
\G_{0,A;0,B} = k \eta_{0,A;0,B} \sp
\eta^{0,A;0,B}\eta_{0,A;0,B} = {\rm dim}\,h
\end{equation}
\begin{equation}
\G_{1,I;1,J} = k \eta_{1,I;1,J} \sp
\eta^{1,I;1,J} \eta_{1,I;1,J} = {\rm dim}\,g/h
\end{equation}
\end{subequations}
and similarly for $\r =3$ and for the permutation-invariant algebras
in \eqref{gsum}.

Then Eq.~\eqref{Delsp} on $g=\so(2n)$ gives the twist-field conformal weights
for our $\Zm_2$ examples
\nl
\ni {\bf $\Pmo$}
\begin{subequations}
\begin{equation}
\hj_{0,\m \n} ( m \geq 0) | 0 \rangle_\s
=\hj_{1,\m} ( m + \srac{1}{2} \geq 0) | 0 \rangle_\s  = 0
\end{equation}
\begin{equation}
\hat \Delta_0 \left( \Pmo \right)
= \frac{ (2n-1)x}{16(x + 2(n-1))}
\end{equation}
\end{subequations}
\skl {\bf $\Cmgo$}
\begin{subequations}
\begin{equation}
\hj_{0,AB} ( m \geq 0) | 0 \rangle_\s
=\hj_{0,IJ} ( m \geq 0) | 0 \rangle_\s=
\hj_{1,AI} ( m + \srac{1}{2} \geq 0) | 0 \rangle_\s = 0
\end{equation}
\begin{equation}
\hat \Delta_0 \left( \frac{\so (2n) }{\mathbb{A}(2n;r)} \right)
=\frac{r(2n-r)}{16} \frac{x}{x + 2(n-1)} \sp
r = n, \ldots , 2n-1
\end{equation}
\begin{equation}
\Cmo = \frac{\so (2n) }{\mathbb{A}(2n;n)} : \qquad
\hat \Delta_0 \left(\Cmo \right) =\frac{n^2 x}{16(x + 2(n-1))}
\end{equation}
\end{subequations}
where $x$ is the invariant level of the affine $\so(2n)$.

For the twisted triality sectors on $g=\so(8)$, Eq.~\eqref{Delsp}
gives \nl
{\bf $\Tmo$} and {\bf $\Tmto$}
\begin{subequations}
\begin{equation}
\hj_{0,A} ( m \geq 0) | 0 \rangle_\s =
\hj_{\pm 1,\a} ( m \pm \srac{1}{3} \geq 0) | 0 \rangle_\s = 0
\end{equation}
\begin{equation}
\label{trcw}
\hat \Delta_0 \left( \Tmo \right)
=\frac{7}{9} \ \frac{x }{x + 6} \sp
\hat \Delta_0 \left( \Tmto \right)
=\frac{10}{9} \ \frac{x  }{x + 6}
\end{equation}
\end{subequations}
where $x$ is the invariant level of the affine $\so(8)$.
Because $\Tm^2$ and $\Tmt^2$ also have order $\r =3$,
Eq.~\eqref{Delsp} tells us that the conformal weights of the scalar
twist-fields of the twisted sectors $\so(8)/\Tm^2$ and $\so(8)/\Tmt^2$
\begin{equation}
\label{trcw2}
\hat \Delta_0 \left(\frac{\so(8)}{\Tm^2}  \right) =
\hat \Delta_0 \left(  \Tmo\right)
\sp
\hat \Delta_0 \left( \frac{\so(8)}{\Tmt^2} \right)=
\hat \Delta_0 \left(  \Tmto\right)
\end{equation}
are the same as those given in \eqref{trcw}. The twisted sectors
$\so(8)/\Tm^2$ and $\so(8)/\Tmt^2$ are further discussed in
Subsec.~\ref{sectri}.

Note that the scalar twist-field conformal weights are different for each of the
$\Zm_2$-type twisted sectors on $\so (2n)$ and for the twisted
triality sectors $\so(8)/\Tm$ and $\so(8)/\Tmt$, even though all the outer
automorphisms on each $g$ are inner-automorphically equivalent.
 This is not surprising
since it is well known that inner-automorphic spectral flow
\cite{Bardakci:1971nb,Freericks:1988zg} affects conformal weights.

For completeness, we also evaluate \eqref{Delsp} to give the scalar
twist-field conformal weights of the twisted sectors of all the
other outer-automorphic orbifolds on simple $g$:
\begin{subequations}
\begin{eqnarray}
\hat \Delta_0 \left( \frac{\su(n)}{\so(n)} \right) &= &
\frac{ (n-1)(n+2)}{32}\frac{x}{x + n} \label{Delsun}\\
\hat \Delta_0 \left( \frac{\su(2n)}{\mathfrak{sp}(n)} \right)
&= & \frac{ (n-1)(2n+1)}{16}\frac{x}{x + 2n}
\end{eqnarray}
\begin{equation}
\hat \Delta_0 \left( \frac{E_6}{F_4} \right) =
\frac{13}{8} \frac{x}{x + 12} \sp
\hat \Delta_0 \left( \frac{E_6}{C_4} \right) =
\frac{21}{8} \frac{x}{x + 8} \ .
\end{equation}
\end{subequations}
Here we have chosen to label the twisted sectors by their
corresponding symmetric spaces $g/h$, and $x$ is the invariant
level of affine $g$. For the charge conjugation  orbifold on $\su(n)$,
the result \eqref{Delsun} was given earlier in Ref.~\cite{Halpern:2002ww}.

Taken with the corresponding results given in
Refs.~\cite{deBoer:2001nw,Halpern:2002ww,Halpern:2002vh}, this
completes the computation of the scalar twist-field conformal
weights in each sector of all the basic WZW orbifolds.

\subsection{The twisted KZ systems \label{kzsec}}

In the general theory of WZW orbifolds
\cite{deBoer:1999na,Halpern:2000vj,deBoer:2001nw,Halpern:2002ww,Halpern:2002vh}
the description is extended to include the
{\it twisted affine primary fields} $\hg = \hgm \hgp$ of sector
$\s$ and their OPEs, e.g.
\begin{equation}
\label{JgOPE}
\hj_\nrm (z,\s) \hgp (\T,w,\s) = \frac{\hgp(\T,w,\s)}{z-w}
\T_\nrm(T,\s) + \Ozw
\end{equation}
where $\hgp (\T,z,\s)$ is the left-mover twisted affine primary
field in twisted representation $\T \equiv \T(T,\s)$. When acting
on the scalar twist-field state $|0 \rangle_\s$ of sector $\s$,
the twisted affine primary fields create the {\it twisted affine primary
states} \cite{Halpern:2002vh} of sector $\s$
\begin{subequations}
\begin{equation}
| \T \rangle_\s =  \lim_{z \rightarrow 0 } \ \hgp (\T,z,\s) z^{\gamma (\T,\s)}
| 0\rangle_\s
\end{equation}
\begin{equation}
\hj_\nrm (m + \srac{n(r)}{\r(\s)} \geq0 )| \T \rangle_\s
= \de_{m+ \srac{n(r)}{\r(\s)},0}| \T \rangle_\s \T_\nrm (T,\s)
\end{equation}
\end{subequations}
where $\gamma (\T,\s)$ is the so-called matrix exponent of the twisted
affine primary field. Moreover, the OPEs of the twisted affine
primary fields lead to the {\it twisted vertex operator equations}
of sector $\s$ and the
 general left- and right-mover {\it twisted KZ
systems} [12-14]
%\cite{deBoer:2001nw,Halpern:2002ww,Halpern:2002vh} 
of the WZW orbifolds.

For twisted sector $\s$ of any WZW orbifold,
the general twisted left-mover  KZ system
\cite{deBoer:2001nw,Halpern:2002ww,Halpern:2002vh}
has the form
\begin{subequations}
\label{tlmkzeq2}
\begin{equation}
\hat A_+ (\T,z,\s) \equiv {}_\s\langle 0|  \hgp(\T^{(1)},z_1,\s) \hgp(\T^{(2)},z_2,\s)
\cdots \hgp(\T^{(N)},z_N,\s) |0 \rangle_\s
\end{equation}
\begin{equation}
\part_\kappa \hat A_+(\T,z,\s) = \hat A_+ (\T,z,\s) \hat W_\kappa (\T,z,\s) \sp
\kappa = 1 \ldots N \sp \s = 0, \ldots ,N_c -1
\end{equation}
\begin{equation}
\label{gwig2}
\hat A_+ (\T,z,\s) \left( \sum_{\r=1}^N \T_{0\m}^{(\r)} (T,\s) \right)
=0 \sp \forall \;\, \m
\end{equation}
\end{subequations}
where the general twisted left-mover connection $\hat W_\k(\T,z,\s)$
is given in Eq.~\eqref{kzctl}. The general Ward identities in
\eqref{gwig2} are associated to the residual symmetry of each sector
$\s$.

Using the data above, the general twisted left-mover KZ system is easily
evaluated for each of our examples: \nl
\ni {\bf $\Pmo$}
\begin{subequations}
\begin{equation}
\hat W_{\k} (\T,z)  =  \frac{2}{2k + Q}
   \left[ \sum_{\r \neq \k} \frac{1}{z_{\k \r} }\left(
 \sum_{\m <\n} \T_{0,\m \n}^{(\r)} \T_{0,\m \n}^{(\k)}
+ \left( \frac{z_{\r}}{z_{\k}} \right)^{\srac{1}{2}}
\sum_{\m} \T_{1,\m}^{(\r)} \T_{-1,\m}^{(\k)} \right)
- \frac{1}{2 z_{\k}} \sum_{\m}
 \T_{1,\m}^{(\k)} \T_{-1,\m}^{(\k)} \right]
\label{kzpm}
\end{equation}
\begin{equation}
\hat A_+ (\T,z,) \left(\sum_{\k =1}^N \T_{0,\m \n}^{(\k)}\right) =
0 \sp \forall \; \m \n \in \so (2n-1)
\end{equation}
\end{subequations}
\skl {\bf $\Cmgo$}
\begin{subequations}
 \begin{eqnarray}
\hat W_{\k} (\T,z) & = & \frac{2}{2k + Q}
   \left[ \sum_{\r \neq \k} \frac{1}{z_{\k \r} }\left(
 \sum_{A <B} \T_{0,AB}^{(\r)} \T_{0,AB}^{(\k)} +
 \sum_{I <J} \T_{0,IJ}^{(\r)} \T_{0,IJ}^{(\k)}
+ \left( \frac{z_{\r}}{z_{\k}} \right)^{\srac{1}{2}}
\sum_{AI} \T_{1,AI}^{(\r)} \T_{-1,AI}^{(\k)} \right) \right.
 \nn \\
 & & \left. \hskip 2cm
- \frac{1}{2 z_{\k}} \sum_{AI}
 \T_{1,AI}^{(\k)} \T_{-1,AI}^{(\k)} \right]
\label{kzcm}
\end{eqnarray}
\begin{equation}
\hat A_+ (\T,z) \left(\sum_{\k =1}^N \T_{0,AB}^{(\k)} \right)=
\hat A_+ (\T,z) \left(\sum_{\k =1}^N \T_{0,IJ}^{(\k)}\right) =
0 \sp \forall \; AB,\ IJ  \in \so (r) \oplus \so (2n-r)
\end{equation}
\end{subequations}
\skl {\bf $\Tmo$}
\begin{subequations}
\label{TmoKZ}
 \begin{eqnarray}
\hat W_{\k} (\T,z) & \!\!=\!\! & \frac{2}{2k + Q}
   \left[ \sum_{\r \neq \k} \frac{1}{z_{\k \r} }\left(
 \sum_{A=1}^{14} \T_{0,A}^{(\r)} \T_{0,A}^{(\k)}
+
\sum_{\a=1}^7 \left(
\left( \frac{z_{\r}}{z_{\k}} \right)^{\srac{1}{3}}
\T_{1,\a}^{(\r)} \T_{-1,\a}^{(\k)} +
 \left( \frac{z_{\r}}{z_{\k}} \right)^{\srac{2}{3}}
 \T_{-1,\a}^{(\r)} \T_{1,\a}^{(\k)} \right) \right) \right. \nn \\
 & & \left. \hskip 2cm
- \frac{1}{3 z_{\k}} \sum_{\a=1}^7 \left(
 \T_{1,\a}^{(\k)} \T_{-1, \a}^{(\k)} + 2
 \T_{-1,\a}^{(\k)} \T_{1, \a}^{(\k)}  \right) \right]
\label{kztm}
\end{eqnarray}
\begin{equation}
\hat A_+ (\T,z)\left( \sum_{\k =1}^N \T_{0,A}^{(\k)} \right)= 0
\sp \forall \; A \in \gt
\end{equation}
\end{subequations}
\skl {\bf $\Tmto$}
\begin{subequations}
\label{TmtoKZ}
 \begin{eqnarray}
\hat W_{\k} (\T,z) & \!\!=\!\! & \frac{2}{2k + Q}
   \left[ \sum_{\r \neq \k} \frac{1}{z_{\k \r} }\left(
 \sum_{A=1}^{8} \T_{0,A}^{(\r)} \T_{0,A}^{(\k)}
+
\sum_{\a=1}^{10} \left(
\left( \frac{z_{\r}}{z_{\k}} \right)^{\srac{1}{3}}
\T_{1,\a}^{(\r)} \T_{-1,\a}^{(\k)} +
 \left( \frac{z_{\r}}{z_{\k}} \right)^{\srac{2}{3}}
 \T_{-1,\a}^{(\r)} \T_{1,\a}^{(\k)} \right) \right) \right. \nn \\
 & & \left. \hskip 2cm
- \frac{1}{3 z_{\k}} \sum_{\a=1}^{10} \left(
 \T_{1,\a}^{(\k)} \T_{-1, \a}^{(\k)} + 2
 \T_{-1,\a}^{(\k)} \T_{1, \a}^{(\k)}  \right) \right]
\label{kztmt}
\end{eqnarray}
\begin{equation}
\hat A_+ (\T,z)\left( \sum_{\k =1}^N \T_{0,A}^{(\k)} \right)= 0
\sp \forall \; A \in \su(3) \ .
\end{equation}
\end{subequations}
The twisted KZ systems of the sectors
$\so (8)/\Tm^2$ and $\so(8)/\Tmt^2$ are discussed in
Subsec.~\ref{sectri}.

The twisted representation matrices $\T = \T(T,\s)$ satisfy
the general {\it orbifold Lie algebra}
\begin{subequations}
\begin{equation}
 [\T_\nrm,\T_\nsn ]  =
i \F_{\nrm, \nsn}{}^{n(r) + n(s),\de} (\s) \T_{n(r)+n(s),\de}
\end{equation}
\begin{equation}
\T_{n(r) \pm \r(\s),\m} = \T_\nrm \sp
[\T^{(\r)}, \T^{(\k)}] = 0 \sp \r \neq \k
\end{equation}
\end{subequations}
in sector $\s$ of the general WZW orbifold, where $\F (\s)$ are
the same twisted structure constants which appear in the general twisted current
algebra \eqref{malg0}. For our examples then,
the orbifold Lie algebra takes the specific forms: \nl
{\bf $\Pmo$}
\begin{subequations}
\begin{equation}
[ \T_{0,\m \n},\T_{0,\r \s} ]
 =   i (\de_{\n \r} \T_{0,\m \s} + \de_{\m \s} \T_{0,\n \r}
  -\de_{\m \r} \T_{0,\n \s} - \de_{\n \s} \T_{0,\m \r})
\end{equation}
\begin{equation}
[ \T_{0,\m \n},\T_{1, \r} ]
 =   i (\de_{\n \r} \T_{1, \m} - \de_{\m \r} \T_{1,\n} ) \sp
[ \T_{1,\m},\T_{1,\n} ] = - i \T_{2,\m\n}
 =   - i \T_{0,\m \n}
\end{equation}
\end{subequations}
\skl {\bf $\Cmgo$}
\begin{subequations}
\begin{equation}
[ \T_{0,AB},\T_{0,CD} ]
 =  i (\de_{BC} \T_{0,AD} + \de_{AD} \T_{0,BC}  -\de_{AC}
 \T_{0,BD} - \de_{BD} \T_{0,AC} )
\end{equation}
\begin{equation}
[ \T_{0,IJ},\T_{0,KL} ]
 =    i (\de_{JK} \T_{0,IL} + \de_{IL} \T_{0,JK}
  -\de_{IK} \T_{0,JL} - \de_{JL} \T_{0,IK} )
\end{equation}
\begin{equation}
[ \T_{0,AB},\T_{0,IJ} ] = 0
\end{equation}
\begin{equation}
[ \T_{1,AI},\T_{1,BJ} ]
 =  - i (\de_{AB} \T_{0,IJ} + \de_{IJ} \T_{0,AB})
\end{equation}
\begin{equation}
[ \T_{0,AB},\T_{1,CI} ]
 =   i (\de_{BC} \T_{1,AI}- \de_{AC} \T_{1,BI} )
\end{equation}
\begin{equation}
[ \T_{0,IJ},\T_{1,AK} ]
 =   i (\de_{JK} \T_{1,AI} - \de_{IK} \T_{1,AJ})
\end{equation}
\end{subequations}
\skl {\bf $\Tmo$}
\begin{subequations}
\label{so8alg}
\begin{equation}
[ \T_{0,A}, \T_{0,B} ] = i f_{ABC} \T_{0,C} \sp
[ \T_{0,A}, \T_{\pm 1, \a} ] = i f_{A\a \be} \T_{\pm 1,\be}
\end{equation}
\begin{equation}
[ \T_{\pm 1,\a}, \T_{\pm 1,\be} ] =\pm \sqrt{\frac{2}{3}}
g_{\a \be \ga} \T_{\pm 2,\ga} =
\pm \sqrt{\frac{2}{3}}
g_{\a \be \ga} \T_{\mp 1,\ga}
\end{equation}
\begin{equation}
\label{TTtr}
[ \T_{\pm 1,\a}, \T_{\mp 1,\be} ] = i f_{\a \be A} \T_{0,A}
\end{equation}
\end{subequations}
\skl {\bf $\Tmto$}
\begin{subequations}
\label{so8alg2}
\begin{equation}
[ \T_{0,A}, \T_{0,B} ] = i f_{ABC} \T_{0,C}
\end{equation}
\begin{equation}
[ \T_{0,A}, \T_{1, \a} ] = -(T_A^{(\irrep{10})})_{\a\be} \T_{+1,\be}
\sp [ \T_{0,A}, \T_{-1, \a} ] =-(\bar T_A^{(\irrep{10})})_{\a\be} \T_{-1,\be}
\end{equation}
\begin{equation}
\label{TpTm}
[ \T_{+1,\a}, \T_{-1,\be} ] =-(T_A^{(\irrep{10})})_{\a\be}
\T_{0,A} \sp
[ \T_{-1,\a}, \T_{+1,\be} ] =-(\bar T_A^{(\irrep{10})})_{\a\be}
\T_{0,A}
\end{equation}
\begin{equation}
[ \T_{\pm 1,\a}, \T_{\pm 1,\be} ] =\pm
\tilde g_{\a \be \ga} \T_{\mp 1,\ga} \ .
\end{equation}
\end{subequations}
More explicit forms of the twisted representation matrices are
discussed in the following subsection.

Using \eqref{TTtr}, \eqref{TpTm} and the fact that
${\rm Tr}(T_A^{(\irrep{10})})=0$, we obtain the simplified form of
the triality connection for $\so(8)/\Tm$ and $\so(8)/\Tmt$
 \begin{eqnarray}
\hat W_{\k} (\T,z) & \!\!=\!\! & \frac{2}{2k + Q}
   \left[ \sum_{\r \neq \k} \frac{1}{z_{\k \r} }\left(
 \sum_{A} \T_{0,A}^{(\r)} \T_{0,A}^{(\k)}
+
\sum_{\a} \left(
\left( \frac{z_{\r}}{z_{\k}} \right)^{\srac{1}{3}}
\T_{1,\a}^{(\r)} \T_{-1,\a}^{(\k)} +
 \left( \frac{z_{\r}}{z_{\k}} \right)^{\srac{2}{3}}
 \T_{-1,\a}^{(\r)} \T_{1,\a}^{(\k)} \right) \right) \right. \nn \\
 & & \left. \hskip 2cm
- \frac{1}{z_{\k}} \sum_{\a} \
 \T_{1,\a}^{(\k)} \T_{-1, \a}^{(\k)}
  \right]
\label{kztm2}
\end{eqnarray}
where the ranges of $A$ and $\a$ for the two cases are given in
\eqref{TmoKZ} and \eqref{TmtoKZ}.
Although flatness of the twisted connections is guaranteed by
the construction, we have used \eqref{so8alg} and \eqref{so8alg2}
to check explicitly and at length
that the twisted triality connection \eqref{kztm2} is not only flat
but abelian flat
\begin{equation}
\partial_\r \hat W_\k (\T,z) -\partial_\k \hat W_\r (\T,z) =
[ \hat W_\k (\T,z), \hat W_\r (\T,z)] = 0
\end{equation}
for both $\so(8)/\Tm$ and $\so(8)/\Tmt$. As noted in Subsec.~\ref{sectri},
the twisted KZ connections of the twisted sectors $\so(8)/\Tmm^2$,
$\Tmm= \Tm$ and $\Tmt$ are also abelian flat.

Including the explicit
check \cite{Halpern:2002ww} of flatness for all orbifolds of
$\Zm_2$-type, this completes the explicit check of flatness
for the twisted KZ connections of each sector of all the
outer-automorphic WZW orbifolds on simple $g$.

\subsection{Representation theory \label{secrep}}

We turn now to discuss the explicit form of the twisted
representation matrices.

For each sector $\s$ of any WZW orbifold $A_g (H)/H$, the general formula for
the twisted representation matrices
\begin{equation}
\T_\nrm \equiv \T_\nrm (T,\s) = \chi_\nrm (\s)
U(\s)_\nrm{}^a U(T,\s) T_a U\hc(T,\s)
\end{equation}
is given in Ref.~\cite{deBoer:2001nw}. Here $T_a ,\,a=1...\text{dim }g$ 
can be any untwisted matrix representation of $g$. The quantities $U\hc(\s)$ 
and $U\hc(T,\s)$ are the eigenvalue matrices of the $H$-eigenvalue problem
and the extended $H$-eigenvalue problem respectively, while
$\{ \chi \}$ is a set of normalization constants. 

Since we are starting in a diagonal basis for each automorphism, we may 
set $\chi (\s) = U\hc(\s)=1$ (see Eq.~\eqref{trivsol}) to obtain
a simplified form for the twisted representation matrices
\begin{equation}
\label{twreq}
 \T(T,\s) = U(T,\s) T U\hc(T,\s) \ .
\end{equation}
Nevertheless, we must still solve the {\it linkage relation}
\cite{deBoer:2001nw} for $W(h_\s;T)$ given $\o (h_\s)$ and the
{\it extended $H$-eigenvalue problem} \cite{deBoer:2001nw}
for $U\hc(T,\s)$
\begin{subequations}
\label{linkrel0}
\begin{equation}
\label{linkrel}
W\hc(h_\s;T) T_a W(h_\s;T) = \o (h_\s)_a{}^b T_b \equiv T_a{}'
\end{equation}
\begin{equation}
\label{ext}
W(h_\s;T) U\hc (T,\s) = U\hc(T,\s) E(T,\s)
\end{equation}
\end{subequations}
in order to evaluate the twisted representation matrices in
\eqref{twreq}. Here $\o (h_\s)$ is the action of $h_\s \in H 
\subset Aut(g)$ on the untwisted currents of affine $g$ and $W(h_\s;T)$ is 
the action of $h_\s$ in untwisted representation $T$.

Solution of the relations in \eqref{linkrel0} is straightforward for
 any ``real'' irrep $T$ of $g$, where real is defined here as the unitary
 equivalence
\begin{equation}
T' = \o T  \cong T
\end{equation}
for any automorphism group $H$ of any $g$.
A simple example is the adjoint representation $T^{\rm adj}$ of any $g$,
for which it is known that \cite{deBoer:2001nw,Halpern:2002ww}
\begin{equation}
(T_a^{\rm adj})_b{}^c = -if_{ab}{}^c \sp
W(h_\s;T^{\rm adj})= \o (h_\s)\sp U(T^{\rm adj},\s) = U\hc(\s)
\end{equation}
in each sector of every WZW orbifold. In our case, this gives the
simple form
\begin{equation}
U\hc(\s) = 1 \qquad \Rightarrow \qquad \T (T^{\rm adj},\s) = T^{\rm adj}
\end{equation}
and the specific results
\begin{equation}
\Pmo \quad : \qquad \T_{0,\m\n}( T^{\rm adj})  = T_{\m \n}^{\rm adj} \sp
\T_{1,\m}( T^{\rm adj}) = T_{\m ,2n}^{\rm adj}
\end{equation}
\begin{equation}
\Cmgo \quad : \qquad \T_{0,AB}( T^{\rm adj}) = T_{AB}^{\rm adj}
\sp\T_{0,IJ} ( T^{\rm adj})= T_{IJ}^{\rm adj}
\sp \T_{1,AI}( T^{\rm adj}) = T_{AI}^{\rm adj}
\end{equation}
\begin{equation}
\Tmo \quad : \quad
\T_{0,A}( T^{\rm adj}) = T_{A}^{\rm adj}= (\r_A)_{\m \n}
T_{\m \n}^{\rm adj} ,\quad
 \T_{\pm 1,\a} ( T^{\rm adj}) \!=
T_\a^{\rm adj}{}^{\pm }= \frac{1}{\sqrt{2}} ( T_{\a 8}^{\rm adj} \pm \frac{i}{2\sqrt{3}}
g_{\a \be \ga} T_{\be \ga}^{\rm adj})
\end{equation}
\begin{equation}
\Tmto \quad : \quad
\T_{0,A}( T^{\rm adj}) = T_{A}^{\rm adj}=  - \frac{1}{2}
f_{Aij} T_{ij}^{\rm adj} \sp
 \T_{\pm 1,\a} ( T^{\rm adj}) =
\frac{1}{4} e^{\pm \srac{\pi i}{6}} (g_{\a}^\pm)_{ij}
 T_{ij }^{\rm adj}
\end{equation}
are obtained for the twisted sectors of this paper. More generally, the
solution of the linkage relation \eqref{linkrel} is guaranteed for
real representations  by the unitary equivalence $T' \cong T$.

For outer automorphisms of simple $g$ we must also consider ``complex''
irreps $T^{(\rm c)}$ for which
\begin{equation}
T^{(\rm c)}{}' = \o  T^{(\rm c)}\ncong T^{(\rm c)}
\end{equation}
and in this case we must take the representation $T$ to be
{\it reducible}
\cite{deBoer:2001nw,Halpern:2002ww}. As an example, for any outer
automorphism $\o^2=1$ of the $\Zm_2$ type it is
known that \cite{Halpern:2002ww}
\begin{subequations}
\begin{equation}
\label{comrep}
T_a =\left( \begin{array}{cc}
T_a^{\rm (c)}& 0 \\ 0 & T_a^{\rm (c)}{}' \end{array} \right)
=\left( \begin{array}{cc}
T_a^{\rm (c)}{} & 0 \\ 0 & \w_a{}^b  T_b^{\rm (c)}{} \end{array} \right)
\end{equation}
\begin{equation}
T_a{}' = \o_a{}^b T_b =\left( \begin{array}{cc}
T_a^{\rm (c)}{}'& 0 \\ 0 & T_a^{\rm (c)}{} \end{array} \right)
\end{equation}
\begin{equation}
W(T) =
i \left( \begin{array}{cc}
0 & \tone \\ \tone & 0
\end{array} \right) \sp W\hc(T) W(T) =\left( \begin{array}{cc}
\tone & 0 \\ 0 & \tone
\end{array} \right)
\end{equation}
\begin{equation}
\label{exHval}
U(T) = U\hc(T) = \frac{1}{\sqrt{2}}
\left( \begin{array}{cc}
\tone & \tone \\ \tone & -\tone
\end{array} \right) \sp U\hc(T) U(T) =\left( \begin{array}{cc}
\tone & 0 \\ 0 & \tone
\end{array} \right) \sp E(T)= i \left( \begin{array}{cc}
\tone & 0 \\ 0 & - \tone
\end{array} \right)
\end{equation}
\end{subequations}
where $W(T)$ and $U(T),E(T)$ are respectively the solutions of
the linkage relation \eqref{linkrel} and the extended $H$-eigenvalue
problem in \eqref{ext}. Then Eq.~\eqref{twreq} gives in particular
for any complex representation
%$\so (2n)/\Pm$ and $\so (2n)/\Cm$:
\nl
\ni {\bf $\Pmo$}
\begin{subequations}
\label{twrepT1}
\begin{equation}
T^{\rm (c)} =(T_{\m \n}^{\rm (c)},T_{\m,2n}^{\rm (c)}) \sp
T^{\rm (c)}{}' = (T_{\m \n}^{\rm (c)},-T_{\m,2n}^{\rm (c)})
\end{equation}
\begin{equation}
\T_{0,\m\n} (T)= U(T) T_{\m\n} U\hc (T) =
\left( \begin{array}{cc}
T_{\m \n}^{\rm (c)} & 0 \\ 0 & T_{\m \n}^{\rm  (c)} \end{array} \right)
\end{equation}
\begin{equation}
\T_{1,\m} (T) = U(T) T_{\m,2n} U\hc(T) =
\left( \begin{array}{cc} 0 & T_{\m,2n}^{\rm (c)} \\ T_{\m,2n}^{\rm (c)} & 0 \end{array} \right)
\end{equation}
\end{subequations}
\skl { $\left\{\frac{\so(2n)}{\Cmgi}, \,r \text{ odd} \right\}$}
\begin{subequations}
\label{twrepT2}
\begin{equation}
T^{\rm (c)} =(T_{AB}^{\rm (c)},T_{IJ}^{\rm (c)},T_{AI}^{\rm (c)}) \sp
T^{\rm (c)}{}' =(T_{AB}^{\rm (c)},T_{IJ}^{\rm (c)},-T_{AI}^{\rm (c)})
\end{equation}
\begin{equation}
\T_{0,AB} (T)= U(T) T_{AB} U\hc (T) =
\left( \begin{array}{cc}
T_{AB}^{\rm (c)} & 0 \\ 0 & T_{AB}^{\rm  (c)} \end{array} \right)
\end{equation}
\begin{equation}
\T_{0,IJ} (T)= U(T) T_{IJ} U\hc (T) =
\left( \begin{array}{cc}
T_{IJ}^{\rm (c)} & 0 \\ 0 & T_{IJ}^{\rm  (c)} \end{array} \right)
\end{equation}
\begin{equation}
\T_{1,AI} (T) = U(T) T_{AI} U\hc(T) =
\left( \begin{array}{cc} 0 & T_{AI}^{\rm (c)} \\ T_{AI}^{\rm (c)} & 0 \end{array} \right)
\end{equation}
\end{subequations}
in each of our outer-automorphically twisted sectors of $\Zm_2$-type. 
Specific examples of Eqs.~\eqref{twrepT1},
\eqref{twrepT2} are obtained by substitution of the Weyl spinor
reps $T^{\rm (c)} = S^{\rm (c)} \equiv S$ in App.~\ref{apprep}.
The conjugate Weyl spinors are automatically included as
$C^{\rm (c)} \equiv S^{\rm (c)}{}'$ in this formulation.

For either of the triality automorphisms $\Tm$ or $\Tmt$
we find instead
\begin{subequations}
\label{triW}
\begin{equation}
T_a =\left( \begin{array}{ccc}
T_a^{\rm (c)}& 0 & 0  \\ 0 & T_a^{\rm (c)}{}' & 0 \\
0 &  0 & T_a^{\rm (c)}{}'{}'
 \end{array} \right)
=\left( \begin{array}{ccc}
T_a^{\rm (c)}{} & 0 & 0 \\
0 & \w_a{}^b  T_b^{\rm (c)} & 0 \\
0 & 0 & (\w^2)_a{}^b T_b^{\rm (c)}
 \end{array} \right)
\end{equation}
\begin{equation}
T_a{}' = \o_a{}^b T_b =\left( \begin{array}{ccc}
T_a^{\rm (c)}{}'& 0 & 0  \\ 0 & T_a^{\rm (c)}{}'' & 0 \\
0 &  0 & T_a^{\rm (c)}
 \end{array} \right)
 \end{equation}
\begin{equation}
W(T) =  \left( \begin{array}{ccc}
0 & 0 & \tone  \\ \tone & 0 & 0 \\ 0 & \tone & 0
\end{array} \right) \sp W\hc(T) W(T) =
W^3 (T) =\left( \begin{array}{ccc}
\tone & 0 & 0 \\ 0 &  \tone & 0 \\ 0 & 0 & \tone
\end{array} \right)
\end{equation}
\begin{equation}
 U\hc(T) = \frac{1}{\sqrt{3}} \left( \begin{array}{ccc}
\tone & e^{2 \pi i /3} \tone  & e^{-2 \pi i/3} \tone \\
\tone & e^{-2 \pi i/3} \tone & e^{2 \pi i/3}\tone \\
\tone & \tone & \tone \end{array} \right)
\sp U\hc(T) U(T) =\left( \begin{array}{ccc}
\tone & 0 & 0 \\ 0 & \tone & 0 \\ 0 & 0  & \tone
\end{array} \right)
\end{equation}
\begin{equation}
 E(T)=  \left( \begin{array}{ccc}
\tone & 0 & 0 \\ 0 & e^{-2\pi i/3}\tone & 0 \\ 0 & 0 &
 e^{2 \pi i/3} \tone
\end{array} \right)
\end{equation}
\end{subequations}
where $W(T)$ and $U\hc(T) ,\,E(T)$ are respectively the solutions of 
the linkage relation \eqref{linkrel} and the extended $H$-eigenvalue 
problem in \eqref{ext}.

To be more explicit for the twisted triality sectors, we introduce
the unified notation
\begin{subequations}
\begin{equation}
T^{(\rm c)} = (T_A^{(\rm c)},T_\a^{(\rm c)}{}^+,T_\a^{(\rm c)}{}^-)
\end{equation}
\begin{equation}
T^{(\rm c)}{}' = (T_A^{(\rm c)},e^{-\srac{2\pi i}{3}}
T_\a^{(\rm c)}{}^+, e^{\srac{2\pi i}{3}}T_\a^{(\rm c)}{}^-)
\sp   T^{(\rm c)}{}'' = (T_A^{(\rm c)},e^{\srac{2\pi i}{3}}
T_\a^{(\rm c)}{}^+,e^{-\srac{2\pi i}{3}}T_\a^{(\rm c)}{}^-)
\end{equation}
\begin{eqnarray}
\Tm \quad & :  &\quad T_A^{(\rm c)} = (\r_A)_{\m\n} T_{\m \n}^{(\rm c)}
\sp T_\a^{(\rm c)}{}^\pm = \frac{1}{\sqrt{2}}
\left( T_{\a 8}^{(\rm c)} \pm \frac{i}{2 \sqrt{3}} g_{\a \be \ga}
T_{\be \ga}^{(\rm c)} \right) \\
\Tmt \quad &  : & \quad T_A^{(\rm c)} = -\frac{1}{2} f_{Aij} T_{ij }^{(\rm c)}
\sp T_\a^{(\rm c)}{}^\pm = \frac{1}{4} e^{\pm \srac{\pi i}{6}} (g_{\a}^\pm)_{ij}
T_{ij }^{(\rm c)}
\end{eqnarray}
\end{subequations}
 where $T^{(\rm c)}$ is any complex irrep of $\so(8) \cong \spi (8)$
 under $\Tm$ or $\Tmt$. Then the twisted representation matrices
 $\T (T)$ of  $\so(8)/\Tm$ or $\so(8)/\Tmt$
\begin{subequations}
\label{triTt}
\begin{equation}
\T_{0,A} (T) = U(T) T_{A} U\hc(T) = \left( \begin{array}{ccc}
T_A^{(\rm c)} & 0 & 0 \\ 0 & T_A^{(\rm c)} & 0 \\ 0 & 0 & T_A^{(\rm c)}
\end{array} \right)
\end{equation}
\begin{equation}
\T_{1,\a}(T) = U(T) T_{\a}^+ U\hc (T) = e^{\srac{2\pi i}{3}}
\left( \begin{array}{ccc}
 0 & T_\a^{(\rm c)}{}^+ & 0  \\ 0 & 0 & T_\a^{(\rm c)}{}^+ \\
T_\a^{(\rm c)}{}^+ & 0 & 0 \end{array} \right)
\end{equation}
\begin{equation}
\T_{-1,\a}(T) = U(T) T_{\a}^- U\hc (T)= e^{-\srac{2\pi i}{3}}
\left( \begin{array}{ccc} 0 & 0 & T_\a^{(\rm c)}{}^-   \\
T_\a^{(\rm c)}{}^- & 0 & 0 \\ 0 & T_\a^{(\rm c)}{}^-  & 0 \end{array} \right)
\end{equation}
\end{subequations}
are obtained from Eqs.~\eqref{twreq} and  \eqref{triW} .

As an explicit example, let us work  out the untwisted and twisted
representation matrices corresponding to the complex reps
$V^{\rm (c)}$, $S^{\rm (c)}$ and $C^{\rm (c)}$ of $\so(8) \cong \spi (8)$
under $\Tm$ or $\Tmt$. We begin with the familiar form of the
untwisted vector rep
$V^{\rm (c)}$ of $\so (8)$ in the standard Cartesian basis:
\begin{subequations}
\label{vscrep}
\begin{equation}
V_{ij}^{\rm (c)} = 2i e_{ij} \sp (e_{ij})_{kl} = \frac{1}{2} (\de_{ik} \de_{jl}
-\de_{jk} \de_{il})
\end{equation}
\begin{eqnarray}
\Tm \quad & :  &\quad V_A^{(\rm c)} = 2 i(\r_A)_{\m\n} e_{\m \n}
\sp V_\a^{(\rm c)}{}^\pm = \frac{2i}{\sqrt{2}}
\left( e_{\a 8} \pm \frac{i}{2 \sqrt{3}} g_{\a \be \ga} e_{\be \ga}
 \right) \\
\Tmt \quad &  : & \quad V_A^{(\rm c)} = -i  f_{Aij} e_{ij }
\sp V_\a^{(\rm c)}{}^\pm = \frac{i}{2} e^{\pm \srac{\pi i}{6}}
(g_{\a}^\pm)_{ij} e_{ij } \ .
\end{eqnarray}
\end{subequations}
In this case, we can define both the untwisted Weyl spinor rep
and the untwisted conjugate Weyl spinor rep directly from the
vector rep and the $\Zm_3$ automorphism
\begin{subequations}
\begin{equation}
S^{(\rm c)} = V^{(\rm c)}{}' = \o V^{(\rm c)}
=(V_A^{(\rm c)}, e^{- \srac{2 \pi i }{3}} V_\a^{(\rm c)}{}^+,
 e^{ \srac{2 \pi i }{3}} V_\a^{(\rm c)}{}^-)
\end{equation}
\begin{equation}
 C^{(\rm c)} = S^{(\rm c)}{}' = \o V^{(\rm c)}{}' = \o^2 V^{(\rm c)}
=(V_A^{(\rm c)}, e^{+ \srac{2 \pi i }{3}} V_\a^{(\rm c)}{}^+,
 e^{ -\srac{2 \pi i }{3}} V_\a^{(\rm c)}{}^-)
\end{equation}
\end{subequations}
where $\o$ can be taken as $\o(\Tm)$ or $\o(\Tmt)$.
Then the required reducible  representation $T$ is
\begin{equation}
T_a
=\left( \begin{array}{ccc}
V_a^{(\rm c)} & 0 & 0 \\ 0 & (\omega V^{(\rm c)})_a & 0 \\
 0 & 0 & (\omega^2 V^{(\rm c)})_a
\end{array} \right)
\end{equation}
and the twisted representation matrices
\begin{subequations}
\label{triTte}
\begin{equation}
\T_{0,A} (T) =\left( \begin{array}{ccc}
V_A^{(\rm c)} & 0 & 0 \\ 0 & V_A^{(\rm c)} & 0 \\
 0 & 0 &  V_A^{(\rm c)}
\end{array} \right)
\end{equation}
\begin{equation}
\T_{+1,\a} (T) = e^{ \srac{2\pi i}{3}}
\left( \begin{array}{ccc}
 0 & V_\a^{(\rm c)}{}^+ & 0  \\ 0 & 0 & V_\a^{(\rm c)}{}^+ \\
V_\a^{(\rm c)}{}^+ & 0 & 0 \end{array} \right)
\sp
\T_{-1,\a} (T) =e^{ -\srac{2\pi i}{3}}
\left( \begin{array}{ccc} 0 & 0 & V_\a^{(\rm c)}{}^-   \\
V_\a^{(\rm c)}{}^- & 0 & 0 \\ 0 & V_\a^{(\rm c)}{}^-  & 0 \end{array} \right)
\end{equation}
\end{subequations}
follow from \eqref{triTt}. For the twisted triality sectors
$\so(8)/\Tm$ and $\so(8)/\Tmt$ the explicit forms of the entries
here are given in Eqs.~(\ref{vscrep}b,c).

\subsection{Action formulation of outer-automorphic WZW orbifolds
 \label{secelt}}

The classical theory of WZW orbifolds is described by the {\it general
 WZW orbifold action} \cite{deBoer:2001nw,Halpern:2002ww,Halpern:2002zv}
on the cylinder $(\xi,t)$ and the solid cylinder $\Gamma$, which reduces to the form
\begin{subequations}
\label{orbact0}
\begin{eqnarray}
\hat S [\hg (\T,\s)]  & = & -\frac{k}{\epsilon y(T)} \left(
\frac{1}{8\pi}\int d^2\xi
 \0b {\rm Tr}\big{(}\;\hat{g}^{-1} (\st,\s)
\pl_+\hat{g}(\st,\s)\0b\hat{g}^{-1}
(\st,\s)\pl_-\hat{g}(\st,\s)\;\big{)} \right. \nn \\
\label{orbact}
 & & \hskip 1.5cm \left.
  +\frac{1}{12\pi}\int_{\Gamma} {\rm Tr}\big{(}\;(\;\hat{g}^{-1}
  (\st,\s) d\hat{g}(\st,\s)\;)^3\,\big{)} \right)
\end{eqnarray}
\begin{equation}
{\rm Tr}(T_a T_b) = y(T) \de_{ab}
\sp \epsilon = \left\{ \begin{array}{ll}
1 & \mbox{for real reps}\,\; T \\
2 & \mbox{for complex reps}\;\, T^{\rm (c)} \;\,\mbox{when}\;\,
\r=2 \\
3 & \mbox{for complex reps}\;\, T^{\rm (c)} \;\,\mbox{when}\;\,
\r=3 \\
\end{array} \right.
\end{equation}
\end{subequations}
for each twisted sector $\s$ of all the
outer-automorphic WZW orbifolds on simple $g$. Here
$\hg (\T,\s) \equiv \hg(\T(T,\s),\xi,t,\s)$ are the
{\it group orbifold elements} of sector $\s$, which are the high-level
or classical limit of the twisted affine primary fields. The
group orbifold elements are locally group elements but they
exhibit the monodromy
\begin{equation}
\label{classmon}
\hg (\T,\xi + 2\pi,t,\s) = E(T,\s) \hg (\T,\xi,t,\s) E(T,\s)^*
\end{equation}
where $E(T,\s)$ is the eigenvalue matrix of the extended
$H$-eigenvalue problem in \eqref{ext}. The result \eqref{orbact0}
generalizes the action given for the charge-conjugation orbifold
on $\su(n)$ in Ref.~\cite{Halpern:2002ww}.

The group orbifold elements can be expressed in terms of the
twisted tangent space coordinates $\hb$
\begin{equation}
\label{tangsp}
\hg (\T(T,\s),\xi,t,\s) = e^{i \hb^\nrm (\xi,t) \T_\nrm (T,\s)}
\sp \hb^\nrm (\xi + 2\pi,t) = \hb^\nrm (\xi,t)
e^{ 2 \pi i \srac{n(r)}{\r (\s)}}
\end{equation}
where $\T_\nrm (T,\s)$ are the same twisted representation matrices
discussed in the operator formulation above. The consistency
of the monodromy of $\hb$ in \eqref{tangsp} and that of $\hg$ in
\eqref{classmon} is a consequence of a selection rule for the twisted
representation matrices \cite{deBoer:2001nw}.

This gives the explicit forms of the group orbifold elements for
each of our twisted sectors \nl
\ni{\bf $\Pmo$}
\begin{equation}
\hg (\T,\xi) = e^{i (\hb^{0,\m\n} (\xi) \T_{0,\m\n} +
\hb^{1,\m} (\xi) \T_{1,\m})} \sp
\hb^{0,\m\n} (\xi +2\pi) = \hb^{0,\m\n} (\xi) \sp
\hb^{1,\m} (\xi + 2\pi )= -\hb^{1,\m} (\xi)
\end{equation}
\skl {\bf $\Cmgo$}
\begin{subequations}
\begin{equation}
\hg (\T,\xi) = e^{i (\hb^{0,AB} (\xi) \T_{0,AB} +
\hb^{0,IJ} (\xi) \T_{0,IJ} + \hb^{1,AI} (\xi) \T_{1,AI})}
\end{equation}
\begin{equation}
\hb^{0,AB} (\xi +2\pi) = \hb^{0,AB} (\xi) \sp
\hb^{0,IJ} (\xi +2\pi) = \hb^{0,IJ} (\xi) \sp
\hb^{1,AI} (\xi + 2\pi )= -\hb^{1,AI} (\xi)
\end{equation}
\end{subequations}
\skl {\bf $\Tmo$} and {\bf $\Tmto$}
\begin{subequations}
\begin{equation}
\hg (\T,\xi) = e^{i (\hb^{0,A} (\xi) \T_{0,A} +
\hb^{\pm 1,\a} (\xi) \T_{\pm 1,\a})}
\end{equation}
\begin{equation}
\hb^{0,A} (\xi+2\pi ) = \hb^{0,A} (\xi) \sp
\hb^{\pm 1,\a} (\xi+ 2 \pi ) =
\hb^{\pm 1,\a} (\xi)\ e^{\pm \srac{2\pi i}{3}}
\end{equation}
\end{subequations}
where $\T = \T(T,\s)$ and we have suppressed the time label $t$.
The explicit forms of $\T$ for real and complex representations
are discussed in the previous subsection. The corresponding
discussion for the charge conjugation orbifold on $\su(n)$ was
given in Ref.~\cite{Halpern:2002ww}.

\section{Assembling the Orbifolds \label{secass}}

\subsection{The $\Zm_2$ orbifolds on $\so (2n)$}

Using the development above, we may construct the orbifolds
\begin{equation}
\frac{A_{\so(2n)}\Big(\Zm_2 (\Pm)\Big)}{\Zm_2 (\Pm)} \sp
\frac{A_{\so(2n)}\Big(\Zm_2 (\Cmgi)\Big)}{\Zm_2 (\Cmgi)}
\sp r = n, \ldots, 2n -2
\end{equation}
of type $\Zm_2$ on $\so (2n)$. These orbifolds have an untwisted
sector $\s=0$ and one twisted sector $\s =1$, called
respectively $\so(2n)/\Pm$ and $\so(2n)/\Cmgi$, $r=n, \ldots, 2n -2$
in this paper. Among these, only those generated by
 $\Pm$ and $\{\Cmgi$, $r={\rm odd}\}$
are outer-automorphic orbifolds, while the others are
inner automorphic. Following the counting in Subsec.~\ref{intsec},
we  have then constructed the number $N$ of distinct $\Zm_2$-type
outer-automorphic orbifolds on $\so(2n\geq 6)$
\begin{equation}
N = \left\{ \begin{array}{cl}
r & \mbox{on}\;\, \so (4r) \\
r+1 & \mbox{on}\;\, \so (4r+2) \ .
\end{array} \right.
\end{equation}
In what follows, we consider the more intricate
orbifolds of types $\Zm_3$ and $S_3$ on $\so (8)$.

\subsection{Two $\Zm_3$ triality orbifolds on $\so(8)$ \label{sectri}}

There are two outer-automorphic $\Zm_3$ orbifolds on $\so(8)$
\begin{equation}
\frac{A_{\so(8)}\Big(\Zm_3 (\Tm)\Big)}{\Zm_3 (\Tm)} \sp
\frac{A_{\so(8)}\Big(\Zm_3 (\Tmt)\Big)}{\Zm_3 (\Tmt)}
\end{equation}
and each of these orbifolds has  three sectors: the untwisted sector $\s=0$, a first
twisted sector $\s=1$ called $\so(8)/\Tm$ or $\so(8)/\Tmt$ above,
and a second twisted sector $\s=2$ which corresponds to
$\so(8)/\Tm^2$ or $\so(8)/\Tmt^2$ respectively. Our task in
this subsection is the description of the $\s=2$ sectors.

For $\Tmm = \Tm$ or $\Tmt$, the action $\omega ( \Tmm^2)
= \omega(\Tmm)^2$ is given by
\begin{equation}
\label{tris}
\Tmm^2 \quad: \qquad
J_A(z)' = J_A(z) \sp J_\a ^\pm(z)' = e^{\pm \srac{2\pi i}{3}}
J_\a^\pm (z)
\end{equation}
in the same diagonal bases given above for $\Tm$ and $\Tmt$.
The untwisted affine-Sugawara construction \eqref{Ttr} holds as
well for $\Tm^2$ and $\Tmt^2$.

The phase reversal for $J_\a^\pm (z)'$ in \eqref{tris} relative to the action of $\omega(\Tmm)$
tells us that the twisted current algebra of
$\so(8)/\Tm^2$ or $\so(8)/\Tmt^2$ is the same as that given for
$\so(8)/\Tm$ or $\so(8)/\Tmt$
in Eqs.~\eqref{tralg} and \eqref{tralg2}, but with
the map
\begin{equation}
\Tmmo \ \rightarrow \ \Tmmos \ : \qquad
m \pm \srac{1}{3} \ \rightarrow \ m \mp \srac{1}{3}
\sp
\hj_{\pm 1,\a} (m \pm \srac{1}{3}) \
\rightarrow \ \hj_{\mp 1,\a} (m \mp \srac{1}{3})
\end{equation}
for both cases.
We may then rewrite the twisted current algebras of $\so(8)/\Tmm^2$
in the standard forms
given for $\so(8)/\Tmm$ in  Eqs.~\eqref{tralg} and \eqref{tralg2}.

For example, we find
\begin{equation}
\Tmos \ : \qquad [ \hj_{\pm 1,\a}(m \pm \srac{1}{3}),
\hj_{\pm 1,\be} (n \pm \srac{1}{3} )] =
\mp \sqrt{\frac{2}{3}}
g_{\a \be \ga}  \hj_{\mp 1,\ga}(m+n \pm 1 \mp \srac{1}{3} )
\end{equation}
instead of \eqref{tralgc} for $\so(8)/\Tm$. For the $\so(8)/\Tm^2$ sector of
the $\Zm_3(\Tm)$ orbifold, this is the only change in the form of
the twisted current algebra.

For the  $\so(8)/\Tmt^2$ sector of
the $\Zm_3(\Tmt)$ orbifold, we find again the standard form
\eqref{tralg2}, but with the replacement
\begin{equation}
\Tmtos \ : \qquad [ \hj_{\pm 1,\a}(m \pm \srac{1}{3}),
\hj_{\pm 1,\be} (n \pm \srac{1}{3} )] =
\mp
\tilde g_{\a \be \ga}  \hj_{\mp 1,\ga}(m+n \pm 1 \mp \srac{1}{3} )
\end{equation}
and $T^{(\irrep{10})} \leftrightarrow \bar T^{(\irrep{10})} $
everywhere.

The rectifications \eqref{recT1} or \eqref{recT2}, the twisted
affine-Sugawara constructions \eqref{tras} and the scalar
twist-field conformal weights \eqref{trcw2} are the same for
the sectors $\so(8)/\Tmm^2$ as they are for the sectors
$\so(8)/\Tmm$.

The twisted representation matrices $\T(T,\s)$ of the $\s=2$ sectors
satisfy the same orbifold Lie algebras \eqref{so8alg} and
\eqref{so8alg2} but with the map
\begin{equation}
\Tmmo \ \rightarrow \ \Tmmos \ : \qquad
\T_{\pm 1,\a} (T) \ \rightarrow \ \T_{\mp 1,\a} (T)
\end{equation}
which mirrors the twisted current algebras. This means
 that we can construct the twisted representation
matrices of sector $\s=2$ from the ones discussed above for
sector $\s=1$:
\begin{equation}
\T_A(T,\s=2) = \T_A (T,\s=1) \sp
\T_{\pm 1,\a} (T,\s=2) = \T_{\mp 1,\a} (T,\s=1) \ .
\end{equation}
Then the explicit forms of the twisted KZ connections and Ward
identities for the
$\s=2$ sectors of the $\Zm_3$ triality orbifolds are
\begin{subequations}
 \begin{eqnarray}
  & & \hskip -1cm  \hat W_{\k} (\T,z,\s=2) \nn \\
  & & =  \frac{2}{2k + Q}
   \left[ \sum_{\r \neq \k} \frac{1}{z_{\k \r} }\left(
 \sum_{A} \T_{0,A}^{(\r)} \T_{0,A}^{(\k)}
+
\sum_\a \left(
\left( \frac{z_{\r}}{z_{\k}} \right)^{\srac{1}{3}}
\T_{-1,\a}^{(\r)} \T_{1,\a}^{(\k)} +
 \left( \frac{z_{\r}}{z_{\k}} \right)^{\srac{2}{3}}
 \T_{1,\a}^{(\r)} \T_{-1,\a}^{(\k)} \right) \right) \right. \nn
 \\
 & & \left. \hskip 2cm
- \frac{1}{z_{\k}} \sum_\a
 \T_{-1,\a}^{(\k)} \T_{1, \a}^{(\k)}  \right]
\label{kztm3}
\end{eqnarray}
\begin{equation}
\hat A_+ (\T,z,\s=2)\left( \sum_{\k =1}^N \T_{0,A}^{(\k)} \right)= 0
\end{equation}
\begin{equation}
A = 1 \ldots 14 \sp \a = 1 \ldots 7 \qquad \mbox{for}\;\,
\frac{\so(8)}{\Tm^2} \qquad ; \qquad
A = 1 \ldots 8 \sp \a = 1 \ldots 10 \qquad \mbox{for}\;\,
\frac{\so(8)}{\Tmt^2}
\end{equation}
\end{subequations}
where these twisted representation matrices $\T$, which satisfy \eqref{so8alg} and \eqref{so8alg2},
are the same ones we constructed for $\so(8)/\Tm$ and
$\so(8)/\Tmt$ in Subsec.~\ref{secrep}. These twisted connections
are also abelian flat.

For the group-orbifold elements discussed in Subsec.~\ref{secelt},
we find instead the forms
\skl {\bf $\frac{\so(8)}{\Tm^2}$} and {\bf $\frac{\so(8)}{\Tmt^2}$}
\begin{subequations}
\begin{equation}
\hg (\T,\xi + 2 \pi,\s=2) = E(T)^* \hg(\T,\xi,\s=2)E(T)
\end{equation}
\begin{equation}
 \hg(\T,\xi,\s=2) =
e^{i (\hb^{0,A} (\xi) \T_{0,A} +
\hb^{\pm 1,\a} (\xi) \T_{\mp 1,\a})}
\end{equation}
\begin{equation}
\hb^{0,A} (\xi+2\pi ) = \hb^{0,A} (\xi) \sp
\hb^{\pm 1,\a} (\xi+ 2 \pi ) =
\hb^{\pm 1,\a} (\xi)\ e^{\pm \srac{2\pi i}{3}}
\end{equation}
\end{subequations}
where $E(T)$ and these twisted representation matrices are
again the objects defined for the $\s=1$ sectors in
Subsec.~\ref{secrep}.

\def\A{{ \mathbb{A}}}
\def\tA{{ \tilde{\A}}}
\def\P{{ \mathbb{P}}}
\def\C{{ \mathbb{C}}}
\def\so{{ \mathfrak{so}}}
\def\su{{ \mathfrak{su}}}
\def\Aso8{{ A_{\so (8)}}}
\def\TT{{ \mathbb{T}}}
\def\Tone{{ {\TT}_1}}
\def\Ttwo{{ {\TT}_2}}
\def\Zint{{ \mathbb{Z}}}
\def\w{{ \omega}}
\def\wP{{ \omega (\P)}}
\def\wA{{ \omega (\A)}}
\def\wtA{{ \omega (\tA)}}
\def\wAtext{{ \omega (\A (8;5))}}
\def\2p3{{ \frac{2\pi i}{3}}}
\def\p3{{ \frac{\pi i}{3}}}
\def\one{{\mathchoice {\rm 1\mskip-4mu l} {\rm 1\mskip-4mu} {\rm 1\mskip-4.5mu l}
{\rm 1\mskip-5mu l}}}
\def\Ad{{ \dot{A}}}
\def\Bd{{ \dot{B}}}
\def\Id{{ \dot{I}}}
\def\Jd{{ \dot{J}}}
\def\ad{{ \dot{\alpha}}}
\def\bd{{ \dot{\beta}}}
\def\md{{ \dot{\mu}}}
\def\nd{{ \dot{\nu}}}
\def\OA{{ \Omega (\A)}}
\def\OtA{{ \Omega (\tA)}}
\def\de{{ \delta}}
\def\nn{{ \nonumber}}

\subsection{Three $S_3$ triality orbifolds on $\so(8)$ }

In the following subsections, we will assemble the three $S_3$
triality orbifolds on $\so(8)$:
\begin{subequations}
\label{S3list}
\begin{equation}
\frac{\Aso8 (S_3 (\P ,\Tone ))}{S_3 (\P ,\Tone )} ,\quad \frac{\Aso8 (S_3 (\A ,\Ttwo ))}{S_3 (\A ,\Ttwo )} ,\quad
   \frac{\Aso8 (S_3 (\tA ,\Tone ))}{S_3 (\tA ,\Tone )}
\end{equation}
\begin{equation}
\label{P=A85}
\P \simeq \A (8;5) \cong \A \cong \tA \,.
\end{equation}
\end{subequations}
In this notation, each $S_3$ is generated by the $\Zint_2$ and the $\Zint_3$ element shown in its
argument.

Some comment will be helpful about the variety of $\Zm_2$ outer
automorphisms in \eqref{P=A85}.
The outer automorphisms $\P$ and $\A (8;5)$, discussed explicitly
in Sec.~\ref{symsec}, are related
by a non-trivial inner automorphism $K$
\begin{subequations}
\begin{equation}
\Pm \;\; : \quad \frac{\so(8)_x}{\so(7)_x} \qquad ; \qquad
\Am (8;5) \;\; : \quad \frac{\so(8)_x}{\so(5)_x \oplus \so(3)_{2x}}
\end{equation}
\begin{equation}
\P \simeq \A (8;5) : \quad \wP = K \wAtext
\end{equation}
\end{subequations}
which changes the dimension of the invariant subalgebra $h$ of $\P$ versus $\A (8;5)$. The
three $\A$-type automorphisms in \eqref{P=A85} are related by the (trivial) inner automorphisms $u$
and $v$
\begin{equation}
\A (8;5) \cong \A \cong \tA \,: \quad \wA =u^\dagger \wAtext u ,\quad \wtA =v^\dagger \wAtext v
\end{equation}
which act as permutations of the $\so (8)$ indices and preserve
$g/h =\so(8)_x/(\so(5)_x \oplus \so(3)_{2x})$. It is known
\cite{Halpern:2000vj,deBoer:2001nw} that conformal weights are 
invariant under such two-sided automorphisms, and that, indeed,
all the twisted tensors of the orbifold can be taken invariant.\footnote{
See e.g. Eq.~(2.28) of Ref.~\cite{Halpern:2000vj}. For our special
case $U(\A (8;5)) = \thickone$, choose $U(\A) =u$ and $U(\tA) = v$,
with the same $E = \o (\A (8;5))$.}
We may therefore consider the twisted sectors $\so (8)/\A$ and 
$\so (8)/\tA$ as identical to the twisted sector $\so (8)/\A 
(8;5)$ given above.

In what follows, we use the diagonal basis
$(J_A ,J^\pm_\alpha )$ discussed for $\Tone$ and $\Ttwo$ in
Subsecs.~\ref{T1sec} and \ref{threeptfive}.

\subsection{The triality orbifold $\Aso8 (S_3 (\P ,\Tone)) /S_3 (\P ,\Tone)$}

To see the first $S_3$ triality orbifold in \eqref{S3list}, we use
Eqs.~\eqref{Jgt}, \eqref{7combo}, \eqref{omP0}
and \eqref{T1aut} to establish the following transformation properties under $\P$ and $\Tone$
\begin{subequations}
\begin{gather}
\wP \left(J_A (z) ,J^\pm_\alpha (z) \right) = \left( J_A (z) ,-J^\mp_\alpha (z) \right) \\
\w (\Tone) \left( J_A (z),J^\pm_\alpha (z) \right) = \left( J_A (z),e^{\mp \2p3} J^\pm_\alpha (z) \right) \\
\w (\P \Tone) \left( J_A (z),J^\pm_\alpha (z) \right) =\left( J_A (z),-e^{\mp \2p3} J^\mp_\alpha (z) \right) \\
\w (\Tone \P) \left( J_A (z),J^\pm_\alpha (z) \right) =\left( J_A (z),-e^{\pm \2p3} J^\mp_\alpha (z) \right) \\
\w (\Tone^2) \left( J_A (z),J^\pm_\alpha (z) \right) =\left( J_A (z),e^{\pm \2p3} J^\pm_\alpha (z) \right) \\
A=1, \ldots ,14 ,\quad \alpha =1,\ldots ,7
\end{gather}
\end{subequations}
where $\w (CD) = \w (C) \w (D)$ and the index $A$ labels the
invariant $\gt$ subalgebra of $\Tone$.
{} From these transformation properties, we verify that
\begin{equation}
\P^2 = \Tone^3 = (\P \Tone )^2 = \one
\end{equation}
so $\P$ and $\Tone$ generate an $S_3$ whose group table is shown
in Table \ref{S3table} and whose conjugacy classes are
\begin{eqnarray}
\quad \quad & \sigma =0: & \one \nn \\
\quad \quad & \sigma =1: & \P ,\P \Tone ,\Tone \P \nn \\
\quad \quad & \sigma =2: & \Tone ,\Tone^2 \,. \label{S3class}
\end{eqnarray}
Then for the $S_3$ triality orbifold
\begin{equation}
\frac{\Aso8 (S_3 (\P ,\Tone ))}{S_3 (\P ,\Tone)}
\end{equation}
we may choose the twisted sectors $\so (8) /\P$ and $\so (8) /\Tone$ given above as the representatives
for sectors $\sigma =1$ and $\sigma =2$ respectively.

\begin{table}[h]
\begin{center}
\begin{tabular}{|c||c|c|c|c|c|} \hline
 & $\Pm$ & $\Pm \Tm$ & $\Tm \Pm $ & $\Tm$ & $\Tm^2$  \\ \hline \hline
$\Pm $ & $\tone$ & $\Tm$ & $\Tm^2$ & $\Pm \Tm$ & $\Tm \Pm $ \\ \hline
$\Pm \Tm$ & $\Tm^2$ & $\tone$ & $\Tm$ & $\Tm \Pm $ & $\Pm$  \\ \hline
 $\Tm \Pm$ & $\Tm$ & $\Tm^2$ & $\tone$ & $\Pm $ & $\Pm \Tm$  \\ \hline
$\Tm$ & $\Tm \Pm$ & $\Pm$ & $\Pm \Tm$ & $\Tm^2$ & $\tone$  \\ \hline
$\Tm^2$ & $\Pm \Tm$ & $\Tm \Pm$ & $\Pm$ & $\tone $ & $\Tm$  \\ \hline
 \end{tabular}
\end{center}
\caption{$S_3$ group table  generated by $\P$ and $\Tone$.
\label{S3table}}
\end{table}

\subsection{The triality orbifold $\Aso8 (S_3 (\A ,\Ttwo)) /S_3 (\A ,\Ttwo)$
\label{s3twosec}}

For the second $S_3$ triality orbifold in \eqref{S3list}, we first define the action of the outer
automorphism $\A$ with $h = \so(5)_x \oplus \so(3)_{2x}$
\begin{subequations}
\begin{gather}
J_{\Ad \Bd} (z)' = J_{\Ad \Bd} (z) ,\quad J_{\Id \Jd} (z)' = J_{\Id \Jd} (z) ,\quad J_{\Ad \Id} (z)' = -J_{\Ad \Id} (z) \\
\Ad ,\Bd =2,5,7 ,\quad \Id ,\Jd =1,3,4,6,8
\end{gather}
\end{subequations}
which is permutation-equivalent to $\A (8;5)$ in \eqref{omCg0}.
A more convenient form of this action is
\begin{subequations}
\begin{gather}
\wA J_{ij} (z) = \OA_{ii'} \OA_{jj'} J_{i'j'} (z) \sp
i,j,i',j' = 1 \ldots 8 \\
\OA_{\Ad \Bd} =\de_{\Ad \Bd} ,\quad \OA_{\Id \Jd} =-\de_{\Id \Jd} ,\quad \OA_{\Ad \Id} =\OA_{\Id \Ad} =0
\end{gather}
\end{subequations}
where $\OA$ is recognized as the action of charge conjugation $\C$ on $\su (3)$ given in 
Ref.~\cite{Halpern:2002ww}. Then it is not difficult to check that
\begin{subequations}
\begin{gather}
(T_A^{\rm adj} )_{kl} \OA_{ki} \OA_{lj} = \OA_{AB} (T_B^{\rm adj} )_{ij} \\
T_i^{(\irrep{3})} {}' \equiv \OA_{ij} T_j^{(\irrep{3})} = \bar{T}_i^{(\irrep{3})} \\
(g^\pm_\alpha )_{kl} \OA_{ki} \OA_{lj} = (g^\mp_\alpha )_{ij}
\end{gather}
\end{subequations}
where $A,B,i,j = 1 \ldots 8$.
Moreover, we may use these relations and Eqs.~\eqref{su3basis},
\eqref{so8su3} to compute the actions
\begin{subequations}
\label{AT2action}
\begin{gather}
\wA \left( J_{\Ad} (z),J_{\Id} (z),J^\pm_\alpha (z) \right) =\left( J_{\Ad} (z),-J_{\Id} (z),e^{\pm \p3}
   J^\mp_\alpha (z) \right) \\
\w (\Ttwo) \left( J_{\Ad} (z),J_{\Id} (z),J^\pm_\alpha (z) \right) =\left( J_{\Ad} (z),J_{\Id} (z),e^{\mp \2p3}
   J^\pm_\alpha (z) \right) \\
\w (\A \Ttwo) \left( J_{\Ad} (z),J_{\Id} (z),J^\pm_\alpha (z) \right) =\left( J_{\Ad} (z),-J_{\Id} (z),
   e^{\mp \p3} J^\mp_\alpha (z) \right) \\
\w (\Ttwo \A) \left( J_{\Ad} (z),J_{\Id} (z),J^\pm_\alpha (z) \right) =\left( J_{\Ad} (z),-J_{\Id} (z),
   -J^\mp_\alpha (z) \right) \\
\w (\Ttwo^2) \left( J_{\Ad} (z),J_{\Id} (z),J^\pm_\alpha (z) \right) =\left( J_{\Ad} (z),J_{\Id} (z),e^{\pm \2p3}
   J^\pm_\alpha (z) \right) \\
\Ad =2,5,7 ,\quad \Id =1,3,4,6,8 ,\quad \alpha =1,\ldots ,10
\end{gather}
\end{subequations}
where $\{ J_A\} = \{ J_{\Ad} ,J_{\Id} \}$ are the currents of the invariant $\su (3)$ subalgebra of $\Ttwo$.
The $\so (3)_{12x} \subset \su (3)_{3x} $ generated by
$\{ J_{\Ad} \}$ is not the $\so(3)_{2x}$ in the invariant subalgebra
of $\Am$.
{} From \eqref{AT2action} we verify that
\begin{equation}
\A^2 = \Ttwo^3 = (\A \Ttwo )^2 = \one
\end{equation}
and the same $S_3$ group table (see Table \ref{S3table})
and conjugacy classes
\eqref{S3class} are obtained with $\P \rightarrow \A$ and $\Tone \rightarrow \Ttwo$. It follows that,
for the $S_3$ triality orbifold
\begin{equation}
\frac{\Aso8 (S_3 (\A ,\Ttwo ))}{S_3 (\A ,\Ttwo )}
\end{equation}
we may choose the twisted sectors $\so (8) /\A (8;5)$ and $\so (8)/\Ttwo$ given above as the
representatives for sectors $\sigma =1$ and 2 respectively.

\subsection{The triality orbifold $\Aso8 (S_3 (\tA ,\Tone)) /S_3 (\tA ,\Tone)$
\label{sectr3}}

Our last $S_3$ triality orbifold is the most intricate of the three.

We begin by defining the action $\wtA$ of the outer automorphism $\tA$
\begin{subequations}
\begin{gather}
J_{\ad \bd} (z)'=J_{\ad \bd} (z) ,\quad J_{\md \nd} (z)'=J_{\md \nd} (z) ,\quad J_{\ad \md} (z)' =
   -J_{\ad \md} (z) \\
J_{\ad 8} (z)'=J_{\ad 8} ,\quad J_{\md 8} (z)'=-J_{\md 8} (z) \\
\ad ,\bd =1,2,3 ,\quad \md ,\nd =4,5,6,7
\end{gather}
\end{subequations}
which (like the automorphism $\Am$) is permutation-equivalent to $\A (8;5)$ in
\eqref{omCg0}. Relative to the notation of Subsec.~\ref{parsec},
we have decomposed the $\so (7)$ vector label $\mu$ as $\{ \mu \} =\{ \ad ,\md \}$. A more convenient
form of this action is:
\begin{subequations}
\begin{gather}
\wtA J_{\mu \nu} (z) = \OtA_{\mu \mu'} \OtA_{\nu \nu'} J_{\mu' \nu'} (z)
\sp \mu, \nu,\mu', \nu' = 1 \ldots 7  \\
\wtA J_{\mu 8} (z) = \OtA_{\mu \mu'} J_{\mu' 8} (z) \\
\OtA_{\ad \bd} =\de_{\ad \bd} ,\quad \OtA_{\md \nd} =-\de_{\md \nd} ,\quad \OtA_{\ad \md} =\OtA_{\md \ad} =0 \,.
\end{gather}
\end{subequations}
Then it is not difficult to check that
\begin{subequations}
\label{AT1Relns}
\begin{gather}
g_{\ad \mu' \nu'} \OtA_{\mu' \mu} \OtA_{\nu' \nu} = g_{\ad \mu \nu} \sp
g_{\md \nu' \rho'} \OtA_{\nu' \nu} \OtA_{\rho' \rho} = -g_{\md \nu \rho} \\
(\rho_{\Ad} )_{\mu' \nu'} \OtA_{\mu' \mu} \OtA_{\nu' \nu} = (\rho_{\Ad} )_{\mu \nu} \sp
(\rho_{\Id} )_{\mu' \nu'} \OtA_{\mu' \mu} \OtA_{\nu' \nu} = -(\rho_{\Id} )_{\mu \nu} \\
\Ad =1,2,5,6,11,12 ,\quad \Id =3,4,7,8,9,10,13,14
\end{gather}
\end{subequations}
where $g_{\alpha \beta \gamma}$ are the octonionic structure constants in
Eq.~\eqref{oct}  and $\{ \rho_A \} =
\{ \rho_{\Ad} ,\rho_{\Id} \}$ is the trace-orthogonal $\gt$ basis given explicitly
in App.~\ref{rhosol}.

Finally, we use the relations \eqref{AT1Relns} together with
Eqs.~\eqref{Jgt}, \eqref{7combo} to compute the
action of $\tA$ in the $\Tone$ basis
\begin{subequations}
\label{AT1Action}
\begin{align}
\wtA \left( J_{\Ad} (z),J_{\Id} (z),J^\pm_{\ad} (z),J^\pm_{\md} (z) \right) &=\left(J_{\Ad} (z),-J_{\Id} (z),
   J^\mp_{\ad} (z),-J^\mp_{\md} (z) \right) \\
\w (\Tone) \left( J_{\Ad} (z),J_{\Id} (z),J^\pm_{\ad} (z),J^\pm_{\md} (z) \right) &=\left( J_{\Ad} (z),
   J_{\Id} (z), e^{\mp \2p3} J^\pm_{\ad} (z), e^{\mp \2p3} J^\pm_{\md} (z) \right) \\
\w (\tA \Tone) \left( J_{\Ad} (z),J_{\Id} (z),J^\pm_{\ad} (z),J^\pm_{\md} (z) \right) &=\left(J_{\Ad} (z),
   -J_{\Id} (z), e^{\mp \2p3} J^\mp_{\ad} (z), -e^{\mp \2p3} J^\mp_{\md} (z) \right) \\
\w (\Tone \tA) \left( J_{\Ad} (z),J_{\Id} (z),J^\pm_{\ad} (z),J^\pm_{\md} (z) \right) &=\!\left(J_{\Ad} (z),
   -J_{\Id} (z), e^{\pm \2p3} J^\mp_{\ad} (z), -e^{\pm \2p3} J^\mp_{\md} (z) \right) \\
\w (\Tone^2) \left( J_{\Ad} (z),J_{\Id} (z),J^\pm_{\ad} (z),J^\pm_{\md} (z) \right) &=\left(J_{\Ad} (z),
   J_{\Id} (z), e^{\pm \2p3} J^\pm_{\ad} (z), e^{\pm \2p3} J^\pm_{\md} (z) \right)
\end{align}
\begin{equation}
\Ad = 1,2,5,6,11,12 ,\,\,\Id =3,4,7,8,9,10,13,14 ,\quad \ad =1,2,3 ,\,\, \md =4,5,6,7 \,.
\end{equation}
\end{subequations}
where $\{ J_A \} = \{ J_{\Ad}, J_{\Id} \}$ are the currents of the invariant $\gt$ subalgebra of
$\Tone$ and we have also decomposed the $\irrep{7}$'s of $\gt$ as
$\{ J^\pm_\alpha \} = \{ J_{\ad}^\pm ,J_{\md}^\pm \}$.
{} From the action \eqref{AT1Action}, we verify that
\begin{equation}
\tA^2 = \Tone^3 = (\tA \Tone)^2 =\one
\end{equation}
and the same $S_3$ group table (see Table \ref{S3table}) and conjugacy classes
\eqref{S3class} are obtained with $\P \rightarrow \tA$. It follows that, for the $S_3$ triality orbifold
\begin{equation}
\frac{ \Aso8 (S_3 (\tA ,\Tone))}{S_3 (\tA ,\Tone)}
\end{equation}
we may choose the twisted sectors $\so (8) /\A (8;5)$ and $\so (8)/\Tone$ given above as the
representatives for sectors $\sigma =1$ and 2 respectively.

In a sense, this third $S_3$ triality orbifold is a surprise. We have
checked that the phases $E_{\a_0}{}' = e^{i \phi_0} E_{\a_0}$ of the
fixed or invariant simple root operators $E_{\alpha_0}$ in the realization of the
 Dynkin automorphisms are as recorded in Fig.~\ref{fig}, where
 the solid double arrows denote the three $S_3$ triality orbifolds we have
 now constructed. The diagram shows clearly that the relative simplicity seen above for the vertical pairings is correlated
with paired trivial or paired non-trivial phases of the fixed simple root
operators. For the third orbifold, however,
we have mixed a trivial phase for the $\Zint_3$ with a non-trivial phase for
 the $\Zint_2$. The phases of the $\Zm_2$ automorphisms are  further
 discussed in App.~\ref{appproof}.

\begin{picture}(250,160)(0,0)
\thicklines
\put(150,0){ % Modify this origin coordinates if you wish to move the right diagram
\begin{picture}(150,160)
\put(36,112){${\mathbb{P}}$}
\put(13,130){${E_{\alpha_0}}'= E_{\alpha_0}$}
\put(35,58){${\mathbb{T}}_1$}
\put(10,40){${E_{\alpha_0}}'= E_{\alpha_0}$}

\put(136,112){${\mathbb{A}}$}
\put(113,130){${E_{\alpha_0}}'= -E_{\alpha_0}$}
\put(135,58){${\mathbb{T}}_2$}
\put(110,40){${E_{\alpha_0}}'= e^{-\frac{2\pi i}{3}}E_{\alpha_0}$}

\put(40,75){\vector(0,1){30}} % left double arrow
\put(40,105){\vector(0,-1){30}} % left double arrow
%\put(40,75){\line(1,2){5}}\put(40,75){\line(-1,2){5}}
%\put(40,105){\line(1,-2){5}}\put(40,105){\line(-1,-2){5}}

\put(140,75){\vector(0,1){30}} % right double arrow
\put(140,105){\vector(0,-1){30}} % right double arrow
%\put(140,75){\line(1,2){5}}\put(140,75){\line(-1,2){5}}
%\put(140,105){\line(1,-2){5}}\put(140,105){\line(-1,-2){5}}

\put(94,90){\vector(2,1){36}}\put(86,86){\vector(-2,-1){36}} % diagonal
%\put(50,68){\vector(2,1){80}}\put(130,108){\vector(-2,-1){80}} % diagonal
\multiput(130,68)(-18,9){5}{\line(-2,1){10}}
\end{picture}} % End of right diagram

\put(50,0){ % Modify these coordinates to change the position of the Dynkin diagram
\begin{picture}(100,150)
\put(40,90){\circle{4}} % middle circle
\put(10,90){\circle{4}} %
\put(60,110){\circle{4}} %
\put(60,70){\circle{4}} %
\thinlines
\put(42,88){\line(1,-1){16}} %
\put(42,92){\line(1,1){16}} %
\put(38,90){\line(-1,0){26}} %
\put(34,100){$\alpha_0$}
\end{picture}}

\put(120,0){ % Modify these coordinates to change the position of the colon
\begin{picture}(5,100)
\put(10,95){\circle*{2}} %
\put(10,85){\circle*{2}} %
\end{picture}}
%%% \end{picture} \\
%%% {\it Fig. 1: The relations among the various $SO(8)$ triality orbifolds.}
%%% \\ \\}\\
%%% If you want a caption uncomment the three lines above and erase the
%next line.
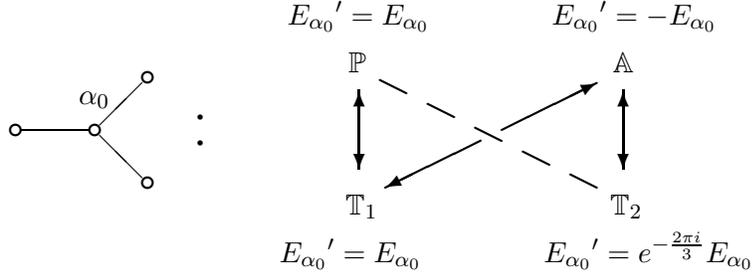
\put(100,10){Fig.\,\ref{fig}: Phases of $E_{\a_0}{}'$ for the $S_3$ triality
orbifolds.}
\end{picture}
\myfig{fig}

The diagram in Fig.~\ref{fig} also suggests the existence of a fourth $S_3$ triality
orbifold corresponding to the dashed line,
but we have proven (see App.~\ref{appproof}) that no such fourth
$S_3$ orbifold can be constructed.

Although all our orbifolds are consistent on the sphere, they should also be checked against
modular invariance on the torus -- but this is beyond the scope of the present paper.

\vskip 1cm
\noindent {\bf Acknowledgements}

We thank D. Bernard, E. Floratos, J. Fuchs, V. Schomerus
and C. Schweigert for helpful discussions.
MBH thanks the Niels Bohr Institute, Jussieu, CERN
and the Max Planck Institut f\"ur
Gravitationsphysik, Golm  for hospitality and support.
The work of the Berkeley authors was supported in part by the Director, Office of Energy Research,
Office of High Energy and Nuclear Physics, Division of High Energy Physics of
the U.S. Department of Energy under Contract DE-AC03-76SF00098 and in part by
the National Science Foundation under grant PHY00-98840.

After submission of this paper, we learned that the twisted current algebra
and twisted affine-Sugawara construction of our sector $\so (8)/\Tm$ had
been discussed earlier in Ref.~\cite{Nep}.

\appendix

\section{Outer automorphisms and invariant subalgebras \label{appinvsub}}

\begin{tabular}{||l|l||} \hline \hline
 & $g/h$   \\ \hline \hline
i & $A_{2r}/B_r$ = $\su (2r+1)/\so (2r+1)$ \\ \hline
ii &  $A_{2r-1}/D_r$ = $\su (2r)/\so (2r)$ \\ \hline
iii & $A_{2r-1}/C_r$ = $\su (2r)/C_r$=$\su (2r)/\mathfrak{sp}(r)$  \\ \hline
iv & $D_{r+1}/B_r$ = $\so (2r+2)/\so (2r+1)$ \\ \hline
v & $D_{r+1}/[B_n \oplus B_{r-n}]$ = $\so (2r+2)/[\so (2n+1)\oplus
  \so (2(r-n)+1)]$ \\ \hline
vi & $E_6/F_4$  \\ \hline
vii & $E_6/C_4$  \\ \hline \hline
viii & $D_4/G_2$ = $\so (8)/\gt $ \\ \hline
ix & $D_4/A_2$ = $\so (8)/\su (3)$ \\ \hline \hline
\end{tabular}
\vskip .5cm
In the form $g/h$, this table gives the inequivalent realizations (homogeneous
gradations) of the Dynkin automorphisms
of simple $g$ in terms of their invariant subalgebras $h$.
To our knowledge this information appeared first in Table 5 of
Ref.~\cite{Bernard:1987rd}.
The first seven entries of the table are of type $\Zm_2$ ($\o^2=1$), where $g/h$
is a symmetric space, while the last two entries are of
type $\Zm_3$ ($\o^3=1$). Different entries for the same $g$ are
inner-automorphically equivalent
(see e.g. Ref.~\cite{Halpern:2002ww}) because each entry
represents the same
 Dynkin automorphism of $g$. Nevertheless, the twisted sector
corresponding to each entry on each $g$ is physically distinct.

The cases i) and ii) complete to $\su (n)/\so (n)$, identified
in Ref.\cite{Halpern:2002ww} as the charge conjugation
automorphism $\Cm$ on $\su (n)$, and the corresponding charge conjugation orbifold
on $\su (n)$ is also discussed in that reference.
The present paper discusses the twisted sectors and orbifolds
corresponding to all the outer automorphisms on $\so(2n)$,
including $\Pm$ (entry iv), $\{\Cmgi, r= {\rm odd} \}$ (entry v), $\Tm$ (entry
viii) and $\Tmt$ (entry ix).

When realizations differ on the same $g$, the difference
is in the phases of the fixed simple root operators under the Dynkin automorphism
(see e.g. Ref.~\cite{Halpern:2002ww}). As an example consider
$\so (2n \geq 6)$ in the Cartan-Weyl basis, with simple
roots and Cartan generators:
\begin{subequations}
\begin{equation}
\a_{(i)} = e_i - e_{i+1} \sp i = 1,\ldots ,n-1 \sp
\a_{(n)} = e_{n-1} + e_n
\end{equation}
\begin{equation}
h_i \equiv e_i \cdot H \sp i = 1 \ldots n \ .
\end{equation}
\end{subequations}
Then we have checked in particular that explicit realizations of the
outer automorphisms $\Pm$ and $\Am \equiv \Am (2n;2n-3)$ may be
completed by starting from the actions on the simple root operators
\begin{subequations}
\begin{equation}
\Pm \left( \frac{\so (2n)}{\so (2n-1)}\right) \ : \qquad
\begin{array}{cc}
E_{\a_{(i)}}{}' = E_{\a_{(i)}} \sp i = 1, \ldots , n-2 \\
%\end{equation}
%\begin{equation}
 E_{\a_{(n-1)}}{}' = E_{\a_{(n)}}
 \sp E_{\a_{(n)}}{}' = E_{\a_{(n-1)}}
 \end{array}
\end{equation}
\begin{equation}
\Am \left( \frac{\so (2n)}{\so (2n-3) \oplus \so(3)}\right) \ : \qquad
\begin{array}{cc}
E_{\a_{(i)}}{}' = E_{\a_{(i)}} \sp i = 1, \ldots , n-3 \sp
 E_{\a_{(n-2)}}{}' = -E_{\a_{(n-2)}}
\\
%\end{equation}
%\begin{equation}
 E_{\a_{(n-1)}}{}' = E_{\a_{(n)}}
 \sp E_{\a_{(n)}}{}' = E_{\a_{(n-1)}}
 \end{array}
\end{equation}
\end{subequations}
and $h_i{}' = h_i$, $i = 1 , \ldots n-1$, $h_n{}' = -h_n$ for
both cases. This computation verifies that the series in entry
v) of the table is outer-automorphic down to and including
$B_1 \equiv \so(3)$.  We also give the generators of the invariant
$B_1 = \so (3) \subset (\so (2n-3) \oplus \so (3)) $ under the
action of $\Am$
\begin{equation}
\{ E_{ \pm \a_{(n-1)}} + E_{ \pm \a_{(n)}} , h_{n-1} \}
\end{equation}
because this case is of special interest in the text
(See Subsecs.~\ref{intsec}, \ref{s3twosec} and \ref{sectr3}).
For the outer automorphisms on $\so(8)$, the phases for the fixed simple
root operators $E_{\a_0}$ are shown in Fig.~\ref{fig}.

\section{Spinor reps of $\mathfrak{spin} (2n)$ \label{apprep}}

We consider as examples some standard \cite{Green:1987sp}
constructions of Dirac and Weyl spinor reps of $\mathfrak{spin} (2n)$.

In the first example, the Dirac gamma matrices $\{\ga_i \}$ are
constructed from $n$ complex fermions as follows
\begin{subequations}
\begin{equation}
\{ b_i ,b_j\hc \} = \de_{ij} \sp
\{ b_i ,b_j \} = \{ b_i\hc ,b_j\hc \} =0 \sp i,j = 1 \ldots n
\end{equation}
\begin{equation}
\ga_i \equiv \left\{
\begin{array}{ll}
b_i + b_i\hc & i = 1 \ldots n \\
\frac{1}{i} (b_{i-n} - b_{i-n}\hc) & i = n+1,\ldots , 2n
\end{array} \right. \sp
\{ \ga_i , \ga_j \}  = 2 \de_{ij} \sp i,j = 1 \ldots 2n
\end{equation}
\begin{equation}
\ga_i = \ga_i\hc =\left(\begin{array}{cc}
0 & \Gamma_i \\
\Gamma_i\hc & 0
\end{array} \right) \sp
\Gamma_i \Gamma_j\hc + \Gamma_j \Gamma_i\hc =
\Gamma_i \hc \Gamma_j + \Gamma_j\hc \Gamma_i = 2 \de_{ij}
\end{equation}
\begin{equation}
\ga_i^* = \ga_i^{\rm t} =\left\{\begin{array}{ll}
\ga_i & i = 1 \ldots n  \\
-\ga_i & i = n+1,\ldots , 2n
\end{array} \right.
\end{equation}
\end{subequations}
where $t$ is matrix transpose.
This gives the Dirac spinor rep $\{D_{ij} \}$ of
$\mathfrak{spin} (2n)$ and its decomposition
into the Weyl reps $S_{ij}$ and $C_{ij}$:
\begin{subequations}
\label{Weyl}
\begin{equation}
D_{ij} =D_{ij}\hc= \frac{i}{4} [ \ga_i , \ga_j] =
\left(\begin{array}{cc}
S_{ij} & 0  \\
0 & C_{ij}
\end{array} \right)
\end{equation}
\begin{equation}
S_{ij} = S_{ij}\hc=\frac{i}{4} (\Gamma_i \Gamma_j\hc - \Gamma_j \Gamma_i\hc )
\sp
%\end{equation}
%\begin{equation}
C_{ij} =C_{ij}\hc= \frac{i}{4} (\Gamma_i \hc \Gamma_j - \Gamma_j\hc \Gamma_i) \ .
\end{equation}
\end{subequations}
Each of these reps satisfy the  algebra of $\so (2n) \cong \mathfrak{spin} (2n)$
\begin{equation}
\label{so2nalg}
[T_{ij}, T_{kl}] = i ( \de_{jk} T_{il} - \de_{ik} T_{jl}
- \de_{jl} T_{ik} + \de_{il} T_{jk} ) \sp T = D, S \;\, \mbox{or}
\;\, C
\end{equation}
as expected.

We know from the text that the automorphic transforms
$T'$ of $T$
\begin{equation}
T'= \o T \sp T = D,S \;\, \mbox{or} \;\, C \sp \o = \o(
\mathbb{A}(2n;r) ) \sp r = n, \ldots , 2n-1
\end{equation}
also satisfy Eq.~\eqref{so2nalg}. For these reps we find that
 $\bar T = T'$ when prime is charge conjugation $\Cm =\mathbb{A}(2n;n) $
\begin{subequations}
\label{Tbarexp}
\begin{eqnarray}
 (\bar T)_{ij} = - (T^{\rm t})_{ij} & =&
\left\{\begin{array}{ll}
T_{ij}  & i,j = 1 \ldots n \;\,\mbox{or} \;\, i,j  =n+1,\ldots , 2n
 \\
-T_{ij} & i = 1 \ldots n ,\ j = n+1,\ldots , 2n
\end{array} \right. \\
&= & T_{ij}{}' \sp T = D,S \;\, \mbox{or} \;\, C
\end{eqnarray}
\end{subequations}
where $\w$ for $\Cm$ is given in \eqref{omC}.
More generally, charge conjugation gives $T' \cong \bar T$ for all
irreps of any Lie $g$. Together, \eqref{inout} and the result
\eqref{Tbarexp} tell us that
\begin{subequations}
\begin{equation}
\spi (4r+2) \sp \Cm \simeq \Pm : \qquad
S_{ij} {}' = \bar S_{ij} \cong C_{ij}
\sp C_{ij} {}' = \bar C_{ij} \cong S_{ij}
\end{equation}
\begin{equation}
\spi (4r) \sp \Cm \not \simeq \Pm : \qquad
S_{ij} {}' = \bar S_{ij} \cong S_{ij}
\sp C_{ij} {}' = \bar C_{ij} \cong C_{ij}
\end{equation}
\end{subequations}
where prime is charge conjugation. In the language of the text,
the Weyl reps are real under $\Cm$ for $\spi (4r)$ and complex
under $\Cm$ for $\spi (4r+2)$.

Other representations can be more natural for other
automorphisms. Consider for example the well-known \cite{Green:1987sp}
Dirac spinor rep of  $\mathfrak{spin} (8)$:
\begin{subequations}
\begin{equation}
\ga_i = \ga_i\hc = \ga_i^{\rm t}  = \left(\begin{array}{cc}
0 & \Gamma_i \\
\Gamma_i\hc & 0
\end{array} \right) \sp
\Gamma_i^* = \Gamma_i \sp
\Gamma_i^{\rm t} =\left\{ \begin{array}{ll}
- \Gamma_i & i = 1 \ldots 7  \\
 \Gamma_8  &  i = 8
\end{array} \right.
\end{equation}
\begin{equation}
\epsilon = i \tau_2 =\left(\begin{array}{cc}
0 & 1 \\
-1 & 0
\end{array} \right)
\end{equation}
\begin{equation}
\Gamma_1 = \epsilon \otimes \epsilon \otimes \epsilon \sp
\Gamma_2 = 1 \otimes \tau_1 \otimes \epsilon \sp
%\end{equation}
%\begin{equation}
\Gamma_3 = 1 \otimes \tau_3 \otimes \epsilon \sp
\Gamma_4 = \tau_1 \otimes \epsilon \otimes 1
\end{equation}
\begin{equation}
\Gamma_5 =  \tau_3 \otimes \epsilon \otimes 1 \sp
\Gamma_6 = - \epsilon \otimes 1 \otimes  \tau_1 \sp
%\end{equation}
%\begin{equation}
\Gamma_7 = \epsilon \otimes 1 \otimes \tau_3 \sp
\Gamma_8 = 1 \otimes 1 \otimes 1
\end{equation}
\end{subequations}
where the Weyl spinor reps $S$ and $C$ are again computed from Eq.~\eqref{Weyl}.
In this case we verify that $S' =C$ and $C' =S$
\begin{subequations}
\begin{eqnarray}
C_{ij} & = &  \frac{i}{4} ( \Gamma_i\hc \Gamma_j -
\Gamma_j\hc \Gamma_i) =
\left\{ \begin{array}{ll}
S_{ij} & i,j = 1 \ldots 7  \\
- S_{i8}  &  j = 8
\end{array} \right. \\
&= & S_{ij} {}' \\
S_{ij} &= & C_{ij} {}'
\end{eqnarray}
\end{subequations}
where prime is the parity automorphism $\Pm$ in \eqref{omP0}.
%So this rep shows very nicely that $\Pm$ is $S \leftrightarrow C$.

\section{More about the embedding $\so (8)_x \supset \so(7)_x
\supset (\gt)_x$ \label{rhosol}}

A trace-orthogonal set of solutions $\{ \r_A, A = 1 \ldots 14 \}$ of
Eq.~\eqref{g2inv} is
\begin{subequations}
\label{rho}
\begin{equation}
\r_1 = \frac{1}{\sqrt{2}} ( e_{23} - e_{67} )
\sp \r_2 = \frac{1}{\sqrt{6}} ( e_{23} + e_{67} - 2 e_{45} ) \sp
\r_3 = \frac{1}{\sqrt{2}} ( e_{37} - e_{14} )
\end{equation}
\begin{equation}
\r_4 = \frac{1}{\sqrt{6}} ( e_{37} + e_{14} + 2 e_{26} ) \sp
\r_5 = \frac{1}{\sqrt{2}} ( e_{47} - e_{56} )
\sp \r_6 = \frac{1}{\sqrt{6}} ( e_{47} + e_{56} + 2 e_{13} ) \sp
\end{equation}
\begin{equation}
\r_7 = \frac{1}{\sqrt{2}} ( e_{16} - e_{24} )
\sp \r_8= \frac{1}{\sqrt{6}} ( e_{16} + e_{24} + 2 e_{35} ) \sp
\r_9 = \frac{1}{\sqrt{2}} ( e_{25} - e_{34} )
\end{equation}
\begin{equation}
\r_{10}=  \frac{1}{\sqrt{6}} ( e_{25} + e_{34} + 2 e_{17} ) \sp
\r_{11} =
\frac{1}{\sqrt{2}} ( e_{12} - e_{46} )
\sp \r_{12} = \frac{1}{\sqrt{6}} ( e_{12} + e_{46} + 2 e_{57} )
\end{equation}
\begin{equation}
\r_{13} = \frac{1}{\sqrt{2}} ( e_{36} - e_{27} )
\sp \r_{14} = \frac{1}{\sqrt{6}} ( e_{36} + e_{27} - 2 e_{15} )
\end{equation}
\end{subequations}
where the antisymmetric tensors $\{ e_{ij} \}$ are defined
in \eqref{eijdef}. Then using $J_A = (\r_A) \cdot J$
and the second part of \eqref{rhogrel}, one obtains the $\gt$ OPE
\eqref{so8gt} and  the following OPEs
\begin{subequations}
\label{so8d0}
\begin{equation}
J_A (z) J_\a(w) =
 \frac{i f_{A\a \be} J_\be (w)}{z-w} + \Ozw \sp
 f_{A\a \be} = - 2 (\r_A)_{\a \be}
\end{equation}
\begin{equation}
J_A (z) J_{\a 8} (w) =
 \frac{i f_{A\a \be} J_{\be 8} (w)}{z-w} + \Ozw
\end{equation}
\begin{equation}
J_{\a 8} (z) J_{\be 8} (w) =\frac{k \de_{\a \be}}{(z-w)^2}
- \frac{i  J_{\a \be} (w)}{z-w} + \Ozw
\end{equation}
\begin{equation}
J_{\a } (z) J_{\be } (w) =\frac{12 k \de_{\a \be}}{(z-w)^2}
+  \frac{2i  (g_{\a \m \ga} g_{\be \ga \n} -g_{\be \m \ga}
g_{\a \ga \n}) J_{\m \n} (w)}{z-w} + \Ozw
\end{equation}
\begin{equation}
J_{\a } (z) J_{\be 8 } (w) = J_{\a8 } (z) J_{\be  } (w) =
  \frac{2 i  g_{\a \be \ga} J_{\ga 8} (w)}{z-w} + \Ozw
\end{equation}
\end{subequations}
 by direct computation. The rest of the OPEs in \eqref{so8d} then
 follow from \eqref{7combo} and \eqref{so8d0}.

In computing \eqref{pmpm} we also find an operator term
proportional to $J_{\m \n}$, but the coefficient is
\begin{equation}
\label{newid1}
6 (e_{\a \be})_{\m \n} - 2 g_{\a \be \ga} g_{\m \n \ga}
- g_{\a \m \ga} g_{\be \n \ga} + g_{\be \m \ga} g_{\a \n \ga}
= 0 \ .
\end{equation}
A consequence of this sum rule is the identity
\begin{equation}
\label{idk}
b_\a d_\be b_\r d_\s g_{\a \be \ga} g_{\r \s \ga} =
b^2 d^2 - (b \cdot d)^2
\end{equation}
for arbitrary 7-vectors $b$ and $d$. The identity \eqref{idk}
allows us to verify the familiar lemma
\begin{equation}
| A B| = |A| |B| = \sqrt{(a^2+b^2)(c^2+d^2)}
\end{equation}
for arbitrary octonions $A$ and $B$ with modulus
$| a + b \cdot i| = \sqrt{a^2 +b^2}$.

Finally, we consider \eqref{pmmp}. In this case, the terms
proportional to $J_{\a 8}$ cancel and the term proportional
to $J_{\m \n}$ has the coefficient
\begin{subequations}
\label{sumr1}
\begin{eqnarray}
 & & - \frac{1}{2}(e_{\a \be})_{\m \n} + \frac{1}{12}
(g_{\a \m \ga} g_{\be \ga \n} - g_{\be \m \ga} g_{\a \ga \n} )  \\
 & & = - (e_{\a \be})_{\m \n} + \frac{1}{6} g_{\a \be \ga}
 g_{\m \n \ga} \label{id2b} \\
 & & = - 2(\r_A)_{\a \be} (\r_A)_{\m \n} \label{id2c}
\end{eqnarray}
\end{subequations}
where we have used the sum rule \eqref{newid1} to obtain
\eqref{id2b}. The last identity in \eqref{id2c} shows that
the operator term is in $\gt$ and verifies that the totally
antisymmetric structure constants satisfy \eqref{fAa} as they
should. Octonionic identities of the type shown in \eqref{newid1}
and \eqref{id2c} are discussed from a different point of view
in Ref.~\cite{Dundarev:1984}.

\newcommand\com[2]{{\lbrack {#1}, {#2} \rbrack}}
\newcommand\Liesu[1]{\mathfrak{su}({#1})}
\newcommand\Lieso[1]{\mathfrak{so}({#1})}
\newcommand\Tth[1]{{T_{#1}^{(\irrep{3})}}}
\newcommand\Tthsb[2]{{\left(\Tth{#1}\right)_{#2}}}
\newcommand\Tadj[1]{{T_{#1}^{\rm adj}}}
\newcommand\Tadjsb[2]{{\left(\Tadj{#1}\right)_{#2}}}
\newcommand\Tten[1]{{T_{#1}^{(\irrep{10})}}}
\newcommand\Ttenb[1]{{\bar T_{#1}^{(\irrep{10})}}}
\newcommand\Ttensb[2]{{\left(\Tten{#1}\right)_{#2}}}
\newcommand\BTten[1]{{\overline{T}_{#1}^{(\irrep{10})}}}
\newcommand\BTtensb[2]{{\left(\BTten{#1}\right)_{#2}}}
\def\Pssu{{\psi_{\Liesu{3}}}}
\def\PssuSq{{\psi_{\Liesu{3}}^2}}
\def\PssuFour{{\psi_{\Liesu{3}}^4}}
\def\lam{{\lambda}}
\newcommand\Tr[1]{{{\mbox{Tr}}\left({#1}\right) }}
\newcommand\gx[2]{{g_{#1}^{#2}}}
\newcommand\gxsb[3]{{\left(\gx{#1}{#2}\right)_{#3}}}
\newcommand\esb[2]{{\left(e_{#1}\right)_{#2}}}
\newcommand\wg[1]{{\widetilde{g}_{#1}}}
\def\Id{{1\kern -.1cm 1}}
\newcommand\Idsb[1]{{{\left(\Id\right)}_{#1}}}

\section{More about the embedding $\Lieso{8}_x\supset \Liesu{3}_{3x}$
\label{appT2}}

We list here some identities which we need to obtain the
diagonal basis of the triality automorphism $\Tmt$ in
Subsec.~\ref{threeptfive}. The basic properties of the $\irrep{3}$
of $\su (3)$ are
\begin{subequations}
\begin{equation}
\Tth{A} = \frac{\sqrt{\PssuSq}}{2}\lam_A \sp
\Tr{\Tth{A}} = 0 \sp  \Tr{\Tth{A}\Tth{B}} = \frac{\PssuSq}{2}\delta_{AB} \sp
A,B =1\ldots 8
\end{equation}
\begin{equation}
\sum_{A=1}^8\Tthsb{A}{IJ}\Tthsb{A}{KL} =
\frac{\PssuSq}{2}(\delta_{IL}\delta_{JK}-\frac{1}{3}
\delta_{IJ}\delta_{KL}) \sp  I,J,K,L=1,2,3
\end{equation}
\end{subequations}
where $\{\lam_A\}$ are the Gell-Mann matrices and $\PssuSq$ is the root length squared
of $\Liesu{3}$.

Using the definitions in the text, we find the identities
\begin{subequations}
\begin{equation}
\left(\Tadj{A}\right)^t = -\Tadj{A} \sp
\left(\gx{\alpha}{\pm}\right)^t = -\gx{\alpha}{\pm}, \quad \alpha=1,\ldots,10
\end{equation}
\begin{equation}
\Tr{T_A^{(\irrep{10})}} = \Tr{\Tadj{A}\gx{\alpha}{\pm}} =
\Tr{\gx{\alpha}{\pm}\gx{\beta}{\pm}} = 0 \sp
\Tr{\gx{\alpha}{\pm}\gx{\beta}{\mp}} = -18\PssuFour
\tone_{\alpha\beta}
\end{equation}
\end{subequations}
where $t$ is matrix transpose. The matrix $\tone$ in \eqref{tone}
is the true identity matrix in the $10\times 10$ space because
\begin{equation}
\tone_{IJK;LMN}B_{LMN} = B_{IJK} \sp \alpha= \{IJK\} =1,\ldots,10
\end{equation}
for any symmetric rank $3$ tensor $B$.

We also find the commutation relations
\begin{subequations}
\begin{eqnarray}
\com{ \Tadj{A} }{ \gx{\alpha}{+} } &=& -\Ttensb{A}{\alpha\beta}\gx{\beta}{+}
\sp
\com{ \Tadj{A} }{ \gx{\alpha}{-} } = -\BTtensb{A}{\alpha\beta}\gx{\beta}{-}
\\
\com{ \gx{\alpha}{+} }{ \gx{\beta}{-} } &=& 6\PssuSq \Ttensb{A}{\alpha\beta}\Tadj{A}
\sp
\com{ \gx{\alpha}{\pm} }{ \gx{\beta}{\pm} } =
-3\PssuSq \wg{\alpha\beta\gamma}\gx{\gamma}{\mp}
\label{Cgpmgpm}
\end{eqnarray}
\end{subequations}
and the tensor $\wg{\alpha\beta\gamma}$ (see \eqref{gtensor})
is fully antisymmetric. Also useful are the sum rules
\begin{subequations}
\begin{equation}
\label{sumr2}
\sum_{\alpha} \left\{
\gxsb{\alpha}{+}{ij}
\gxsb{\alpha}{-}{kl} +
\gxsb{\alpha}{-}{ij}
\gxsb{\alpha}{+}{kl}
\right\} =
18 \PssuFour \esb{ij}{kl}
+6\PssuSq\sum_A\Tadjsb{A}{ij}\Tadjsb{A}{kl}
\end{equation}
\begin{equation}
 \wg{\alpha\gamma\delta}  \wg{\gamma\delta\beta}
 = \sum_A ( \Tten{A}\Tten{A})_{\alpha\beta}
 =\sum_A ( \Ttenb{A}\Ttenb{A})_{\alpha\beta}\ .
\end{equation}
\end{subequations}
In the text, the identities of this appendix are used with $\PssuSq = \frac{2}{3}$.

\def\tP{{ \tilde{\P}}}
\def\wtP{{ \omega (\tP)}}

\section{Nonexistence of a fourth $S_3$ triality orbifold on $\so (8)$
\label{appproof}}

In Sec.~\ref{secass}, we constructed three $S_3$ triality orbifolds
on $\so (8)$ and in Subsec.~\ref{sectr3} we mentioned
the possibility of a fourth $S_3$ triality orbifold on $\so (8)$ based on any $\tP \cong \P$ and the triality
automorphism $\Ttwo$. We now present a proof that no such $S_3$ can be constructed.

We begin by noting some properties of the $\Zint_2$ outer automorphism $\tP$: Since it is a $\Zint_2$ automorphism,
all of the eigenvalues of $\wtP$ are $\pm 1$. Furthermore, we know that any outer automorphism $\tP \cong \P$ on
$\so (8)$ will leave an $\so (7)$ subalgebra invariant. Therefore, $\wtP$ has eigenvalue $+1$ with multiplicity $21$,
and eigenvalue $-1$ with multiplicity $7$. As we shall see, it is this fact which makes it impossible to generate an $S_3$
from $\tP$ and $\Ttwo$.

To understand this, let us assume the opposite, namely that it is possible to generate an $S_3$ from $\tP$ and $\Ttwo$.
Then we have the following necessary and sufficient identities:
\begin{subequations}
\begin{gather}
\Ttwo^3 = \tP^2 = (\tP \Ttwo )^2 = \one \\
\Rightarrow \wtP^2 = \one ,\quad \wtP \w (\Ttwo) = \w (\Ttwo)^2 \wtP \,. \label{wtP-Relns}
\end{gather}
\end{subequations}
In the $\Ttwo$ basis developed in Subsec.~$3.5$, we know the explicit, diagonal form of $\w (\Ttwo)$
\begin{equation}
\label{wT2-form}
\w (\Ttwo) = \left( \begin{array}{ccc} \thickone_8 & 0 & 0\\ 0 & \eta \thickone_{10} & 0\\ 0 & 0 &
   \eta^2 \thickone_{10} \end{array} \right) ,\quad \eta = e^{-\frac{2\pi i}{3}}
\end{equation}
where the diagonal blocks correspond to the action on the currents $(J_A (z), J^+_\alpha (z), J^-_\alpha (z))$
respectively. Using Eq.~\eqref{wT2-form}, we find that the general solution $\wtP$ of the relations \eqref{wtP-Relns} is
\begin{equation}
\wtP = \left( \begin{array}{ccc} A_8 & 0 & 0\\ 0 & 0 & B_{10} \\ 0 & C_{10} & 0 \end{array} \right) ,\quad A_8^2 =\one_8
   ,\quad B_{10} C_{10} = \one_{10}
\end{equation}
and so the $20 \times 20$ submatrix $\tilde{B}$ satisfies
\begin{equation}
\tilde{B} \equiv \left( \begin{array}{cc} 0 & B_{10} \\ C_{10} & 0 \end{array} \right)
,\quad \tilde{B}^2 = \one ,\quad {\rm Tr} (\tilde{B}) = 0 \,.
\end{equation}
It follows that $\tilde{B}$ has 10 eigenvalues $+1$ and 10 eigenvalues $-1$, and hence $\wtP$ (if it exists) has {\it at least
10 eigenvalues} $-1$. We have now reached a contradiction because, as we explained above, $\wtP$ must have exactly 7
eigenvalues $-1$. It follows that there is no fourth $S_3$ triality orbifold on $\so (8)$.

We note in passing that the other $\Zint_2$-type automorphism, namely $\wA$, passes the $S_3$ test above as it must:
Because $\wA$ leaves $\so (3) \oplus \so (5)$ invariant, it has 13 eigenvalues $+1$ and 15 eigenvalues $-1$.

\vskip .5cm
%\clearpage
\addcontentsline{toc}{section}{References}

\renewcommand{\baselinestretch}{.4}\rm
{\footnotesize

%\bibliographystyle{utphys}
%\bibliography{bibliocft}
%\bibliographystyle{../INPUT/utphys}
%\bibliography{../BIB/bibliocft}

%\providecommand{\href}[2]{#2}\begingroup\raggedright

%\endgroup

\providecommand{\href}[2]{#2}\begingroup\raggedright\endgroup

}
\end{document}